\documentclass[journal, 12pt, onecolumn]{IEEEtran}
\usepackage[T1]{fontenc}
\usepackage[latin9]{inputenc}
\usepackage{color}
\usepackage{array}
\usepackage{float}
\usepackage{textcomp}
\usepackage{mathtools}
\usepackage{amsmath}
\usepackage{amssymb}
\usepackage{stackrel}
\usepackage{graphicx}
\usepackage{setspace}
\usepackage[numbers]{natbib}
\usepackage[unicode=true,
 bookmarks=true,bookmarksnumbered=true,bookmarksopen=true,bookmarksopenlevel=1,
 breaklinks=false,pdfborder={0 0 0},pdfborderstyle={},backref=false,colorlinks=false]
 {hyperref}
\hypersetup{pdftitle={Your Title},
 pdfauthor={Your Name},
 pdfpagelayout=OneColumn, pdfnewwindow=true, pdfstartview=XYZ, plainpages=false}
\usepackage[dot]{bibtopic}

\makeatletter

\providecommand{\tabularnewline}{\\}
\floatstyle{ruled}
\newfloat{algorithm}{tbp}{loa}
\providecommand{\algorithmname}{Algorithm}
\floatname{algorithm}{\protect\algorithmname}

\let\oldforeign@language\foreign@language
\DeclareRobustCommand{\foreign@language}[1]{%
  \lowercase{\oldforeign@language{#1}}}

\usepackage[caption=false,font=footnotesize]{subfig}

\makeatother

\begin{document}

\title{Secrecy Rate Maximization for Hardware Impaired \\
Untrusted Relaying Network with Deep Learning }

\author{Hamed~Bastami,~Majid~Moradikia,~ Hamid~Behroozi,~\IEEEmembership{Member,~IEEE,}\\
~Rodrigo~C.~de~Lamare,~\IEEEmembership{Senior~Member,~IEEE,~}Ahmed~Abdelhadi,~\IEEEmembership{Senior~Member,~IEEE,~}\\
and~Zhigou~Ding,~\IEEEmembership{Fellow,~IEEE}\thanks{H.Bastami, H.Behroozi, and Majid Moradikia are with the Department
of Electrical Engineering, Sharif University of Technology, Tehran,
Iran, e-mails: hamed.bastami.ee.sharif.edu, behroozi@sharif.edu, m.moradikia@sutech.ac.ir.}\thanks{RC de Lamare is with the Telecommunications Research (CETUC) PUC-Rio
and University of York, e-mail: \protect\href{mailto:delamare@cetuc.puc-rio.br}{delamare@cetuc.puc-rio.br}.}\thanks{Ahmed Abdelhadi is with the Hume Center for National Security and
Technology, Virginia Tech, Arlington, VA, 22203 USA, e-mail: \protect\href{mailto:aabdelhadi@vt.edu}{aabdelhadi@vt.edu}.}\thanks{Zhiguo Ding is with School of Electrical and Electronic Engineering,
The University of Manchester, Manchester, M13 9PL, UK, e-mail: \protect\href{mailto:zhiguo.ding@manchester.ac.uk}{zhiguo.ding@manchester.ac.uk}.}}

\markboth{}{Hamed Bastami \MakeLowercase{\emph{et al.}}: Secrecy Rate Maximization
for Hardware Impaired Untrusted Relaying Network with Deep Learning }
\maketitle
\begin{abstract}
\textcolor{black}{This paper investigates the physical layer security
design of an untrusted relaying network where the source node coexists
with a multi-antenna eavesdropper (Eve). While the communication relies
on untrustworthy relay nodes to increase reliability, we aim to protect
the confidentiality of information against combined eavesdropping
attacks performed by both untrusted relay nodes and Eve. Taking into
account the hardware impairments, and power budget constraints, this
paper presents a novel approach to jointly optimize relay beamformer
and transmit powers aimed at maximizing average secrecy rate (ASR).
The resultant optimization problem is non-convex, and a suboptimal
solution is obtained through the sequential parametric convex approximation
(SPCA) method. In order to prevent any failure due to infeasibility,
we propose an iterative initialization algorithm to find the feasible
initial point of the original problem. To satisfy low-latency as one
of the main key performance indicators (KPI) required in beyond 5G
(B5G) communications, a computationally efficient data-driven approach
is developed exploiting a deep learning model to improve the ASR while
the computational burden is significantly reduced. Simulation results
assess the effect of different system parameters on the ASR performance
as well as the effectiveness of the proposed deep learning solution
in large-scale cases.}
\end{abstract}


\IEEEpeerreviewmaketitle{}

\section{Introduction}

\IEEEPARstart{I}{n} the last decades, physical layer security (PLS)
has attracted a significant interest as a promising paradigm for establishing
secure communication against unfriendly eavesdroppers (Eves). Unlike
the conventional sophisticated cryptographic techniques, implemented
in the upper layers, PLS provides the secure communication by intelligently
exploiting the time varying nature of fading channels \citep{mukherjee2014principles}.
In this regard, Wyner showed as long as the channel condition of Eve
is a degraded version of the intended receiver, we can hinder the
Eve to overhear the information while the intended receiver can correctly
decode the information from the received signal \citep{wyner1975wire}.
As a result, relay-assisted communication has recently gained much
interest \citep{guo2017joint}-\citep{mashdour2020secure} as an effective
PLS solution. In order to take advantages of multiple intermediate
helpers, i.e., cooperative beamforming (CB) and cooperative jamming
(CJ), have been extensively proposed in the literature \citep{guo2017joint}-\citep{mashdour2020secure}.
The main idea of CB is to adjust the weights at distributed relay
nodes for focusing a narrow beam on the legitimate receiver \citep{yang2013cooperative},
\citep{8334689} or nulling out the relayed information at potential
Eve, i.e., null space beamforming (NSB) \citep{guo2017joint}-\citep{wang2015joint1},
\citep{moradikia2018joint}-\citep{ouyang2017destination}. In contrast,
CJ helps to confuse Eves by isotropically radiating interference artificial
noise (AN) signal \citep{guo2017joint}-\citep{wang2015joint1}, \citep{moradikia2018joint}-\citep{8334689}.
Besides cooperative relay nodes, the usage of destination-aided CJ
(DACJ) can lead to the legitimate destination playing the role of
a jammer \citep{ouyang2017destination}-\citep{8334689}. Notably,
when the CJ and CB techniques are incorporated together, enhanced
secrecy is achieved \citep{guo2017joint}-\citep{wang2015joint1},
\citep{moradikia2018joint}-\citep{ouyang2017destination}. In particular,
DACJ can generally achieve a higher secrecy rate than that obtained
by deploying one or multiple relay nodes to emit jamming noises \citep{ouyang2017destination}.
In fact, by applying DACJ, we can take full advantage of the relay
nodes to amplify and forward (AF) the source signal rather than sacrificing
a relay node and reducing array gain. 

In some practical scenarios such as internet of things (IoT) networks,
a curious node can collaborate as a helper node. Although this untrustworthy
relay node is deployed to improve the communication reliability, it
may attempt to wiretap the information which should be taken into
account as a potential passive attack \citep{moradikia2018joint},
\citep{ouyang2017destination}, \citep{mashdour2020secure}. To address
the security concern of an untrusted relaying network in the presence
of a single-antenna external Eve, a joint CB and CJ design was presented
in \citep{moradikia2018joint} to the secrecy rate maximization. However,
the authors of \citep{moradikia2018joint} considered the ideal hardware,
which is not realistic in practice. The impact of the inherent imperfection
namely hardware impairment (HI), as a more realistic scenario was
taken into account in \citep{8805054}, \citep{8334689}. In particular,
in \citep{8805054}, a different joint CB-CJ scheme was proposed.
Actually, given the common assumption of unknown Eve's CSI \citep{8805054},
\citep{ouyang2017destination}, \citep{8334689}, \citep{mashdour2020secure},
the secrecy rate maximization problem does not work anymore and therefore
it is replaced by another tractable optimization problem in which
the power assigned for information transmission is minimized such
that it needs to be sufficient for satisfying the minimum required
quality of service (QoS) at the destination. Hence the remainder power
budget can be allocated to maximize the jamming sources in order to
enhance the security of the system. This design leads to a sub-optimal
but adequate solution. 

In this paper, we assume perfect CSI at the transmitter (CSIT) and
the receiver (CSIR).\footnote{To practice, there are many practical scenarios where the Eve' CSI
is available. This assumption corresponds to the scenarios where Eve
is one of the network's users, but has not been authorized to receive
the current services offered by the source \citep{wang2013hybrid},
\citep{wang2015joint1}, \citep{moradikia2018joint}. Furthermore,
even for a passive Eve, due to the local oscillator power inadvertently
leaked from the receiver RF front-end, Eve\textquoteright s CSI can
be estimated \citep{6288501}.} It should be noted that the assumption of perfect CSI knowledge is
ideal, however, the result obtained in this paper provides the performance
limit for practical HI relaying communication systems. Given these
cases, despite the adequate PLS design proposed by \citep{8805054},
the maximized secrecy rate cannot be achieved. Additionally, regarding
the solution proposed in \citep{moradikia2018joint}, although the
authors presumed the external Eve, having known CSI, they considered
unrealistic assumptions of sum power constraint at the relays together
with perfect hardware. Moreover, because of the single antenna Eve,
considered in \citep{moradikia2018joint}, the proposed design cannot
be generalized to more difficult scenarios, in the sense of security,
where the Eve has been equipped with multi-antenna.

This paper goes beyond the two abovementioned studies of \citep{moradikia2018joint}
and \citep{8805054} by investigating a joint CB-CJ design with the
goal of secrecy rate maximization. More explicitly, by considering
realistic HIs in an untrusted cooperative network in the presence
of a multi-antenna external Eve, the relay beamformer and transmit
powers are jointly designed so that the achievable secrecy rate (ASR)
is maximized. For this problem, both the total power budget constraint
for the whole network and individual power constraint at each node,
are considered. To combat the combined eavesdropping attacks by untrusted
nodes and external Eve, the DACJ is selected for the first phase and
the idle source node is firstly configured to serve as the jammer
for the second phase. As the formulated optimization problem is non-linear
and non-convex, that is NP-hard, the alternative solution is to deploy
NSB at the cooperative relay nodes to cancel out the information leakage
at Eve, resulting in a simpler non-convex optimization problem. Thanks
to NSB the second cooperative phase is secured and we no longer need
to employ jamming and thus the source node, which was earlier activated
to serve as a jammer in the second phase, might remain silent during
this phase. To solve the corresponding optimization problem we resort
to the sequential parametric convex approximation (SPCA) methodology
\citep{10.2307/169728}, \citep{Beck2010}, leading to a tradeoff
between computational complexity and optimality of the solution. SPCA
results in an iterative algorithm wherein the non-convex feasible
set is suitably approximated by a convex feasible set at each iteration.
Using this convex approximation, a sequence of convex programs can
be efficiently solved, instead. In order to prevent any failure due
 to infeasibility we also develop an iterative initialization algorithm
to find the feasible initial point of the original problem instead
of an arbitrary point, as in the conventional SPCA \citep{10.2307/169728},
\citep{Beck2010}. Despite the excellent results, the computational
complexity order of numerical SPCA-based solution is significantly
increased upon increasing the network dimension including the number
of users, relay nodes, and number of antennas equipped at each node.
This will be verified both analytically and through simulation results
in this paper. 

\textcolor{black}{Satisfying low-latency requirement, as one of the
main Key Performance Indicators (KPIs) of beyond 5G (B5G) \citep{dizdar2020ratesplitting}
communications, would be specifically challenging in face of high
computational load due to large-scale scenarios. To circumvent this
issue, artificial intelligence (AI)-enabled communications comes into
prominence. A subset of key-enabling technologies of AI, so-called
deep learning (DL), is key due to its  nonlinear modeling ability
to solve complicated problems. To be specific, the stronger computing
rate and lower price of deep neural networks (DNN) make it more practical
in wider scenarios, e.g., system performance analysis \citep{8892492}
and wireless resource management \citep{8444648}-\citep{8335785}.
Given the capability of DNN in reducing the computational cost, we
have developed a DL-based approach to solve the non-convex optimization
problem. In the proposed DL approach, the DNN model is firstly trained
to extract a relationship between the system parameters and optimization
variables. Using this relationship, the ASR performance can be efficiently
predicted. }

\textcolor{black}{In summary, the main contributions of this paper
are:}
\begin{itemize}
\item \textcolor{black}{Taking into account the HIs, and both the total
power budget constraint for the whole network and individual power
constraint at each node, }we propose\textcolor{black}{{} a novel approach
to jointly optimizing the relay beamformer and transmit powers aimed
at maximizing the ASR. }While to safeguard the first transmission
phase the DACJ strategy is opted, the security of the second phase
is guaranteed by deploying NSB at the relay nodes. 
\item To maximize the ASR, an iterative algorithm is proposed through solving
the non-convex optimization problem iteratively. We use the SPCA method
to obtain a suboptimal solution. Morever, to avoid any failure due
to infeasibility, an initialization algorithm is proposed to find
the feasible initial point of the original problem. 
\item Despite the similar scenario in \citep{8805054} with unknown Eve's
CSI assumption, our numerical results show that its performance is
significantly lower compared to the performance of the proposed approach.
This is because the proposed approach directly maximizes the ASR and
considerably outperforms the approach in \citep{8805054}. Therefore,
our proposed scheme can be regarded as a upper bound though some relaxation
is adopted to overcome the nonconvexity. 
\item Due to unknown Eve's CSI assumption in \citep{8805054}, a relay selection
algorithm, namely hybrid assisted cooperative jamming (HACJ), which
selects the jammer node among relay nodes and destination, is needed
to improve the secrecy. However, we can get rid of the computational
load imposed by relay selection using simple DACJ, achieving better
performance than HACJ. This is because, assuming perfect CSIT and
CSIR, instead of choosing a relay node as a jammer the entire potential
of relay nodes are preferred to forming a centralized beam towards
the legitimate destination whilst completely nulling out the leakage
at Eve. 
\item To deal with high computational load imposed by large-scale scenarios,
a deep learning-based approach is also proposed where a DNN model
is trained to extract a relationship between the system parameters
and optimization variables. Using this relationship, the ASR performance
can be efficiently predicted and maximized. This model can be generalized
to various optimization problems, leading to SPCA-based solution.
\end{itemize}
\textcolor{black}{The rest of this paper is organized as follows.
The system model is provided in Section II. Section III presents the
problem formulation and the corresponding optimization problem. Section
IV obtain the proposed SPCA-based solution and converts the non-convex
problem into a convex problem. The feasible initialization procedure
is provided in Section V. The deep learning scheme with the aim of
reducing the computational cost in large-scale scenarios is presented
in section VI. Complexity analysis of the proposed scheme is evaluated
in Section VII. In Section VIII, simulation results are presented
to show the effectiveness of the proposed method, and finally the
paper is concluded in Section IX.}

\textit{Notation:} Vectors and matrices are denoted by lower-case
and upper-case boldface symbols, respectively. $\left(.\right)^{\mathrm{T}}$,
$\left(.\right)^{\mathrm{*}}$, $\left(.\right)^{\mathrm{H}}$, and
$\left(.\right)^{\mathrm{-1}}$ denote the transpose, conjugate, conjugate
transpose, and inverse of a matrix respectively. $\mathfrak{R}e(.)$
denote the real part of a complex variable, and $\mathfrak{I}m(.)$
denote the imaginary part of a complex variable. We use $\mathbb{E}{\left\{ \cdot\right\} }$
and $\triangleq$ to denote the expectation and definition operations,
respectively.
\begin{figure}[tbh]
\centering{}\includegraphics[scale=0.46]{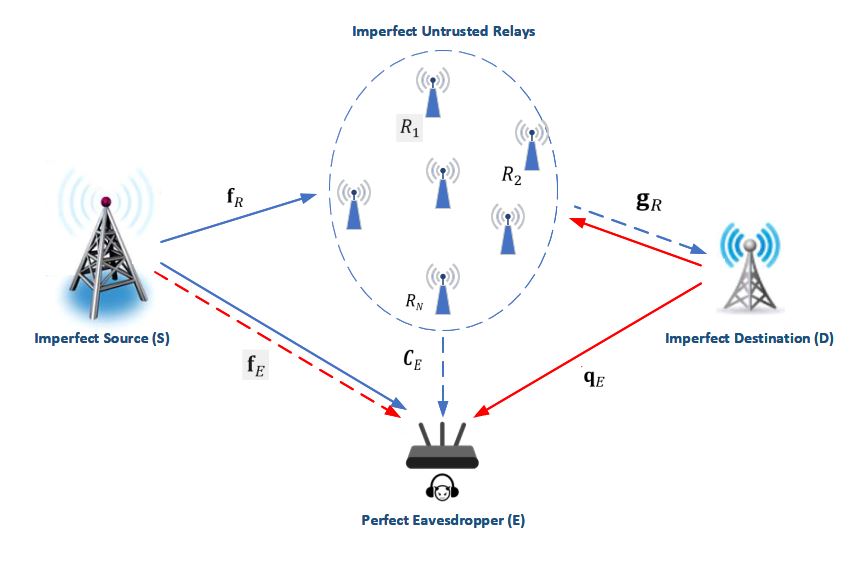}\caption{The considered cooperative network.}
\end{figure}
 A complex Gaussian random variable with mean $\mathit{\mu}$ and
variance $\sigma^{2}$ reads as $\mathcal{C}\mathcal{N}\left(\mathit{\mu},\sigma^{2}\right)$.
Notation $\mathrm{Vec}(\mathbf{H})$ convert matrix $\mathbf{H}$
in single column vector, and $\mathbf{I}_{N}$ denotes $N\times N$
identity matrix. Also, $\mathbb{R}^{N\text{\texttimes}1}$ and $\mathbb{C}^{N\times1}$
denote the set of real and complex $N$-dimentional vector, repectively.
$\mathbb{C}^{N\times N}$ stands for an $N\times N$ complex matrix. 

\section{System model}

A wireless network is considered, as shown in Fig.1, in which a source
node ($\mathrm{\mathit{S}}$) aims to convey the confidential message
to a legitimate destination ($\mathrm{\mathit{D}}$) with the aid
of $\mathit{N}$ distributed intermediate relay node available in
the set of $\mathrm{\mathbf{R}\triangleq{\left\{ \mathit{R}_{i}\right\} }_{i=1}^{\mathit{N}}}$,
in the presence of an external eavesdropper $\mathit{E}$. The relay
nodes in our network are semi-trusted \citep{moradikia2018joint},
i.e., they are trusted to send the accurate CSIs to $\mathit{S}$
via relay\textquoteright s cooperation while they are untrustworthy
for retransmitting the confidential information. It is noteworthy
that the term \textquotedblleft $\mathit{curious\:node}$\textquotedblright{}
throughout the paper comprises both the untrusted relay nodes and
external Eve. We further assume that the untrusted relays are deployed
in non-colluding settings. Besides, given that all nodes are subjected
to the half-duplex constraint and also there is no direct connection
between $\mathit{S}$ and $\mathit{D}$, data transmission takes place
in two consecutive time-slots. We suppose that $\mathit{E}$ is equipped
with $\mathit{N_{E}}$ antenna while all the other nodes have a single
antenna. In our study, all the channels are modelled as block-fading
with channel reciprocity \citep{8227729}. In phase $\mathrm{I}$,
while $\mathit{S}$ broadcasts the information signal with power $\mathit{0<P_{s}\le P_{T}}$,
the destination node ($\mathit{D}$) is configured to send the interference
signal with power $\mathit{0<P_{J_{\textrm{1}}}\leq\bar{P}_{J_{\textrm{1}}}}$
with the aim of confusing the curious nodes. During the second phase
of transmission, called relaying phase, the untrusted amplify-and-forward
(AF) relays within set $\mathbf{R}$ adopt the distributed beamforming
to forward the received signal towards the legitimate terminal $\mathit{D}$.
The power consumed by $\mathbf{R}$ is represented by the vector $\mathbf{P}_{R}\triangleq[P_{R_{1}},P_{R_{2}},\ldots,P_{R_{N}}]^{T}\mathrm{\in}\mathbb{R}^{N\times1}$,
and $\mathit{P_{R_{l}}}$ therein denotes the power consumption at
$\mathit{l}$-th relay node $\mathit{R_{l},\:\forall\:l\in\mathcal{L}}$
with $\mathit{\mathcal{L}\triangleq{\left\{ \mathrm{1},\mathrm{2},\ldots,N\right\} }}$.
Notably, in the relaying phase, although the untrusted relays cannot
decipher the information, they provide another chance for $\mathit{E}$
to extract the information. As such, we should secure the system only
against the Eve in this phase. Hence, in the relaying phase in order
to combat the eavesdropper\textquoteright s attack, we deploy the
source node ($\mathit{S}$) for injecting jamming signal with power
$\mathit{0<P_{J_{\textrm{2}}}\leq\bar{P}_{J_{\textrm{2}}}}.$

The statistical behaviour of HI at node $\mathit{i\mathrm{\in}{\left\{ S,\mathbf{R},D\right\} }}$
is characterized by exploiting the generalized system model of \citep{6630485}.
Note that the ideal hardware eavesdropper is considered in this paper
which implies the worst-case condition in terms of secrecy performance.
Accordingly, denoting the impairment levels at Tx and Rx segments
respectively by the so-called error vector magnitude (EVM)\footnote{The EVM, can be measured directly in practice \citep{book}, e.g.,
3GPP LTE has EVM requirements in the range of $k_{i}^{t}$, $k_{i}^{r}$$\in\left[0.08,0.175\right]$. } parameters $\mathit{k_{i}^{t}}$ and $\mathit{k_{i}^{r}}$, defined
as the ratio of the average distortion magnitude to the average of
signal magnitude, the distortion noise appeared at each node are presented
as: \textbf{\textit{i.}} ${{\eta}_{s}^{t}}\sim\mathcal{C}\mathcal{N}(0,P_{s}{k_{s}^{t}}^{2}),$
\textbf{\textit{ii.}} ${{\eta}_{J_{1}}^{t}}\sim\mathcal{C}\mathcal{N}(0,P_{J_{1}}{k_{J_{1}}^{t}}^{2}),$\textbf{\textit{
iii.}} ${{\eta}_{J_{2}}^{t}}\sim\mathcal{C}\mathcal{N}(0,P_{J_{2}}{k_{J_{2}}^{t}}^{2}),$
\textbf{\textit{iv.}} ${\eta}_{D}^{r}\sim\mathcal{C}\mathcal{N}(0,{k_{D}^{r}}^{2}\sum_{i=1}^{N}{P_{R_{i}}{\left|g_{R_{i}}\right|}^{2}}),$
\textbf{\textit{v.}} ${\boldsymbol{\mathrm{\mathbf{\eta}}}}_{R}^{r}\sim\mathcal{C}\mathcal{N}\left(0,{k_{R}^{r}}^{2}\boldsymbol{\mathrm{\Pi}}\left(P_{s},P_{J_{1}}\right)\right),$
\textbf{\textit{vi.}} ${\boldsymbol{\mathrm{\mathbf{\eta}}}}_{R}^{t}\sim\mathcal{C}\mathcal{N}\left(0,{k_{R}^{t}}^{2}\boldsymbol{\mathrm{\Lambda}}\left(\mathbf{P}_{R}\right)\right),$
where: $\boldsymbol{\mathrm{\Lambda}}\left(\mathbf{P}_{R}\right)\boldsymbol{\triangleq}\mathrm{diag}(\mathbf{P}_{R}),$
and 
\begin{gather*}
\boldsymbol{\mathrm{\Pi}}\left(P_{s},P_{J_{1}}\right)\boldsymbol{\triangleq\,}\mathrm{diag}\left[\text{\ensuremath{\left(P_{s}{\left|f_{R1}\right|}^{2}+{P_{J_{1}}\left|g_{R1}\right|}^{2}\right)}},\right.\\
\left.\dots,\left(P_{s}{\left|f_{R_{N}}\right|}^{2}+P_{J_{1}}{\left|g_{R_{N}}\right|}^{2}\right)\right],
\end{gather*}
\textcolor{black}{where ${\boldsymbol{\mathrm{f}}}_{R}\boldsymbol{\triangleq}{\left[f_{R,\mathrm{1}},\ f_{R,\mathrm{2}},\ \dots,\ f_{R,N}\right]}^{T}$
and ${\boldsymbol{\mathrm{g}}}_{R}\boldsymbol{\triangleq}{\left[g_{R,\mathrm{1}},\ g_{R,\mathrm{2}},\ \dots,\ g_{R,N}\right]}^{T}$
represent the complex-valued channel coefficients from the $S\to\mathbf{R}$
and $\mathbf{R}\to D$, respectively.}

\subsection{Signal Representation}

As discussed above, during the first phase, while $\mathit{S}$ broadcasts
the unit power information symbol $\mathit{x_{s}}$, i.e., $\mathit{\mathbb{E}{\left\{ |x_{s}|^{\mathrm{2}}\right\} }=\mathrm{1}}$,
the destination ($\mathit{D}$) radiates the normalized power jamming
signal $\mathit{z_{1}}$, i.e., $\mathit{\mathbb{E}{\left\{ |z_{1}|^{\mathrm{2}}\right\} }=\mathrm{1}}$,
to cover the information transmission. The signals received at untrusted
relay nodes $\boldsymbol{\mathbf{R}}$, i.e., $\mathit{{\mathbf{\boldsymbol{\boldsymbol{\mathbf{y}}}}}_{R}\mathrm{\in}\mathbb{C}^{N\times1}}$,
and $\mathit{E}$ during phase I, i.e., $\mathit{\boldsymbol{\mathbf{y}}_{E}^{(1)}\in\mathbb{C}^{N_{E}\times1}}$
can be represented by :
\begin{gather}
{\mathbf{\boldsymbol{\boldsymbol{\mathbf{y}}}}}_{R}=\left(\sqrt{P_{s}}x_{s}+{{\eta}_{s}^{t}}\right){\mathbf{f}}_{R}+\left(\sqrt{P_{J_{1}}}z_{1}+{{\eta}_{J_{1}}^{t}}\right){\mathbf{g}}_{R}+{\mathbf{\boldsymbol{\eta}}}_{R}^{r}+\mathbf{n}_{R},\label{eq:7}
\end{gather}
\begin{gather}
\boldsymbol{\mathbf{y}}_{E}^{(1)}=\left(\sqrt{P_{s}}x_{s}+{{\eta}_{s}^{t}}\right){\boldsymbol{\boldsymbol{\mathrm{f}}}}_{E}+\left(\sqrt{P_{J_{1}}}z_{1}+{{\eta}_{J_{1}}^{t}}\right){\boldsymbol{\boldsymbol{\mathrm{q}}}}_{E}+\mathbf{n}_{E}^{(1)},\label{eq:8}
\end{gather}
where ${\boldsymbol{\mathrm{f}}}_{E}\boldsymbol{\in}{\mathbb{C}}^{N_{E}\times1}$and
${\boldsymbol{\mathrm{q}}}_{E}\boldsymbol{\in}{\mathbb{C}}^{N_{E}\times1}$
denote the complex-valued channel coefficients \textcolor{black}{of
the $S\to E$ an}d $D\to E$ links, respectively. $\mathit{\mathbf{n}_{R}\mathrm{\in}\mathbb{C}^{N\times1}}$
and $\mathit{\mathbf{n}_{E}^{(1)}\mathrm{\in}\mathbb{C}^{N_{E}\times1}}$
respectively denote the additive noise at the relay nodes and $\mathit{E}$
in phase I. In relaying phase, the received signal at $\mathbf{R}$
is amplified and forwarded towards $\mathit{D}$. The transmitted
signal by $\mathbf{R}$ is $\mathit{\mathbf{x}_{R}=\mathbf{W}^{H}\mathbf{y}_{R}}$,
in which the weight matrix $\mathbf{W}$ is in the form of $\mathit{\mathbf{W}\triangleq\mathrm{diag}{\left\{ {\boldsymbol{w}}\right\} }}$
with the normalized vector $\mathit{{\boldsymbol{w}}\triangleq[w_{1},w_{2},\ldots,w_{N}]^{T}\mathrm{\in}\mathbb{C}^{N\times1}}$
and $\mathit{w_{l}}$, $\mathit{\forall\,l\mathrm{\in}\mathcal{L}}$,
therein indicates the beamforming weight adopted by the $\mathit{l}$-th
relay. In the meantime, as mentioned before, $\mathit{S}$ jams through
the unit power jamming signal $z_{2}$. Accordingly, the received
signals at $\mathit{D}$, which decodes the information by self-interference
cancellation, and $\mathit{E}$ becomes :
\begin{equation}
y_{\mathrm{D}}={\boldsymbol{\mathrm{g}}}_{R}^{T}\left({\boldsymbol{\mathrm{\mathbf{x}}}}_{R}+{\boldsymbol{\mathrm{\eta}}}_{R}^{t}\right)+{\eta}_{D}^{r}+n_{D}=\sqrt{P_{s}}{{\boldsymbol{\mathrm{g}}}_{R}^{T}{\boldsymbol{\mathrm{W}}}^{H}{\boldsymbol{\mathrm{f}}}_{R}x}_{s}+{\overline{n}}_{D},\label{eq:12}
\end{equation}
\[
{\boldsymbol{\mathrm{y}}}_{E}^{(2)}={\boldsymbol{\mathrm{C}}}_{E}\left({\boldsymbol{\mathrm{x}}}_{R}+{\boldsymbol{\mathrm{\mathrm{\eta}}}}_{R}^{t}\right)+\left(\sqrt{P_{J_{2}}}z_{2}+{{\eta}_{J_{2}}^{t}}\right){\boldsymbol{\mathrm{f}}}_{E}+{\boldsymbol{\mathrm{n}}}_{E}^{(2)}
\]
\begin{gather}
=\sqrt{P_{s}}{{\boldsymbol{\mathrm{C}}}_{E}{\boldsymbol{\mathrm{W}}}^{H}{\boldsymbol{\mathrm{f}}}_{R}x}_{s}+\sqrt{P_{J_{1}}}{\boldsymbol{\mathrm{C}}}_{E}{\boldsymbol{\mathrm{W}}}^{H}{\boldsymbol{\mathrm{g}}}_{R}z_{1}+\sqrt{P_{J_{2}}}\mathbf{f}_{E}z_{2}+{\overline{\boldsymbol{\mathrm{n}}}}_{E}^{(2)},\label{eq:13}
\end{gather}
where the $\mathit{l}$-th column of matrix $\mathit{\mathbf{C}_{E}\mathrm{\in}\mathbb{C}^{N_{E}\times N}}$,
i.e., $\mathit{\mathbf{C}_{E,l}\mathrm{\in}\mathbb{C}^{N_{E}\times1}}$,
denotes the complex-valued channel coefficient from $\mathit{R_{l}\rightarrow E}$,
$\mathit{\forall\,l\mathrm{\in}\mathcal{L}}$. ${\overline{n}}_{D}\boldsymbol{\triangleq}{\boldsymbol{\mathrm{g}}}_{R}^{T}{\boldsymbol{\mathrm{W}}}^{H}{\boldsymbol{\mathrm{f}}}_{R}{{\eta}_{s}^{t}}+{\boldsymbol{\mathrm{g}}}_{R}^{T}{\boldsymbol{\mathrm{W}}}^{H}{\boldsymbol{\mathrm{g}}}_{R}{{\eta}_{J_{1}}^{t}}+{\boldsymbol{\mathrm{g}}}_{R}^{T}{\boldsymbol{\mathrm{W}}}^{H}{\boldsymbol{\mathrm{\eta}}}_{R}^{r}+{\boldsymbol{\mathrm{g}}}_{R}^{T}{\boldsymbol{\mathrm{W}}}^{H}{\boldsymbol{\mathrm{n}}}_{R}+{\boldsymbol{\mathrm{g}}}_{R}^{T}{\boldsymbol{\mathrm{\eta}}}_{R}^{t}+{\eta}_{D}^{r}+n_{D}$,
and ${\overline{\boldsymbol{\mathrm{n}}}}_{E}^{(2)}\boldsymbol{\triangleq}{\boldsymbol{\mathrm{C}}}_{E}{\boldsymbol{\mathrm{W}}}^{H}{\boldsymbol{\mathrm{f}}}_{R}{{\eta}_{s}^{t}}+{\boldsymbol{\mathrm{C}}}_{E}{\boldsymbol{\mathrm{W}}}^{H}{\boldsymbol{\mathrm{g}}}_{R}{{\eta}_{J_{1}}^{t}}+{{\eta}_{J_{2}}^{t}}{\boldsymbol{\mathrm{f}}}_{E}+{\boldsymbol{\mathrm{C}}}_{E}{\boldsymbol{\mathrm{W}}}^{H}{\boldsymbol{\mathrm{\eta}}}_{R}^{r}+{\boldsymbol{\mathrm{C}}}_{E}{\boldsymbol{\mathrm{W}}}^{H}{\boldsymbol{\mathrm{n}}}_{R}+{\boldsymbol{\mathrm{C}}}_{E}{\boldsymbol{\mathrm{\eta}}}_{R}^{t}+{\boldsymbol{\mathrm{n}}}_{E}^{(2)}$and
$\mathit{n_{D}}$ and $\mathit{{\boldsymbol{\mathrm{n}}}_{E}^{\mathrm{(2)}}}$$\mathit{\mathrm{\in}\mathbb{C}^{N_{E}\times1}}$represent
the additive noise at $\mathit{D}$ and $\mathit{E}$, during the
relaying phase, respectively. We normalize the power of $\mathit{z_{\mathrm{2}}}$,
as well, i.e., $\mathit{\mathbb{E}{\left\{ |z_{\mathrm{2}}|^{\mathrm{2}}\right\} }=\mathrm{1}}$
\citep{moradikia2018joint}. 

Note that, $\mathit{\mathbf{x}_{R}}$ have to meet not only the individual
power constraint at each relay no\textcolor{black}{de \citep{wang2015joint1},
\citep{7042285} but }also the total power constraint of the whole
netwo\textcolor{black}{rk }given by:
\begin{equation}
P_{R_{l}}=\mathbb{E}\left\{ {\left|x_{R,l}\right|}^{2}\right\} \le Q_{l},\forall l\boldsymbol{\in}\mathcal{L},\label{eq:10}
\end{equation}
\begin{equation}
P_{tot}=P_{R,tot}+P_{s}+P_{J_{1}}+P_{J_{2}}\le Q_{tot},\label{eq:11}
\end{equation}
where $\mathit{P_{R,tot}=\mathbb{E}\left\{ {\boldsymbol{\mathrm{x}}}_{R}^{H}{\boldsymbol{\mathrm{x}}}_{R}\right\} =\sum_{l=1}^{N}{\mathbb{E}\left\{ {\left|x_{R,l}\right|}^{2}\right\} }}$
stands for the power consumed by $\mathbf{R}$ to retransmit the signal.
$\mathit{Q_{l}}$ denotes the transmit power budget of the $\mathit{l}$-th
relay node, and $\mathit{Q_{tot}}$ is the total power constraint
of the whole network. Given that $\mathit{\boldsymbol{a}^{T}.diag(\boldsymbol{b}^{*})=\boldsymbol{b}^{H}.\mathrm{diag}(\boldsymbol{a})}$,
by defining $\mathit{\mathbf{G}_{R}\triangleq\mathrm{diag}(\mathbf{g}_{R})}$,
one can rewrite Eq. (3) as:

\begin{equation}
y_{\mathrm{D}}=\sqrt{P_{s}}{{\boldsymbol{w}}^{H}{\boldsymbol{\mathrm{G}}}_{R}{\boldsymbol{\mathrm{f}}}_{R}x}_{s}+{\bar{n}}_{D},\label{eq:14-1}
\end{equation}

The equivalent model for the legitimate node $\mathit{D}$ is a SISO
system as shown in (7), whilst $\mathit{E}$ exploits the information
received from both phases to extract the information by incorporating
the observations of two phases, as it was obtained in (2) and (4),
as follows :
\begin{equation}
{\boldsymbol{\mathrm{y}}}_{\mathrm{E}}={\boldsymbol{\mathrm{H}}}_{E}x_{s}+{\boldsymbol{\mathrm{n}}}_{E},\label{eq:15-2}
\end{equation}
where we have:
\begin{equation}
{\boldsymbol{\mathrm{H}}}_{E}=\left[\begin{array}{c}
\sqrt{P_{s}}{\boldsymbol{\mathrm{f}}}_{E}\\
\sqrt{P_{s}}{\boldsymbol{\mathrm{C}}}_{E}{\boldsymbol{\mathrm{F}}}_{R}{\boldsymbol{w}}^{\boldsymbol{*}}
\end{array}\right],\,\,{\boldsymbol{\mathrm{n}}}_{E}=\left[\begin{array}{c}
{\overline{\boldsymbol{\mathrm{n}}}}_{E}^{(1)}\\
\sqrt{P_{J_{1}}}{\boldsymbol{\mathrm{C}}}_{E}{\boldsymbol{\mathrm{G}}}_{R}{\boldsymbol{w}}^{\boldsymbol{*}}z_{1}+\sqrt{P_{J_{2}}}\mathbf{f}_{E}+{\overline{\boldsymbol{\mathrm{n}}}}_{E}^{(2)}
\end{array}\right],\label{eq:16-1}
\end{equation}
with ${\boldsymbol{\mathrm{F}}}_{R}\boldsymbol{\triangleq}\mathrm{diag}({\boldsymbol{\mathrm{f}}}_{R})$
and ${\overline{\boldsymbol{\mathrm{n}}}}_{E}^{(\mathrm{1})}\boldsymbol{\triangleq}\sqrt{P_{J_{1}}}{\boldsymbol{\mathrm{q}}}_{E}z_{1}+{\boldsymbol{\mathrm{f}}}_{E}{{\eta}_{s}^{t}}+{\boldsymbol{\mathrm{q}}}_{E}{{\eta}_{J_{1}}^{t}}+{\boldsymbol{\mathrm{n}}}_{E}^{(1)}$.
In addition, $\mathit{\mathbf{n}_{E}}$ is zero-mean Gaussian vector
with covariance matrix $\mathit{{\boldsymbol{\mathrm{Q}}}_{E}\boldsymbol{=}\mathbb{E}\left\{ {\boldsymbol{\mathrm{n}}}_{E}{\boldsymbol{\mathrm{n}}}_{E}^{H}\right\} \boldsymbol{\in}{\mathbb{C}}^{\mathrm{2}N_{E}\times\mathrm{2}N_{E}}}$.
All the local noise terms $\mathit{n_{D}}$, $\mathit{{\boldsymbol{\mathrm{n}}}_{E}^{(\mathrm{1})}}$,$\mathit{{\overline{\boldsymbol{\mathrm{n}}}}_{E}^{(\mathrm{2})}}$
and $\mathit{\mathbf{n}_{R}}$ are zero-mean and independent complex
Gaussian random variables ($\mathit{r.v.s}$) with variance $\mathit{\sigma^{\mathrm{2}}}$.
Both the jamming signals $\mathit{z_{\mathrm{1}}}$ and $\mathit{z_{\mathrm{2}}}$
are assumed to be complex Gaussian $\mathit{r.v.s}$, as well. 

The end-to-end information rate $\mathit{I(y_{D};x_{s})}$ achieved
by the legitimate terminal is given by:
\begin{gather}
I\left(y_{D};x_{s}\right)=\frac{1}{2}{\mathrm{log}}_{\mathrm{2}}\left(1+\left.\frac{P_{s}{\boldsymbol{w}}^{H}{\boldsymbol{\mathrm{\Phi}}}_{Gf}\boldsymbol{w}}{{\boldsymbol{w}}^{H}{\boldsymbol{\Psi}}_{k}\left(P_{s},P_{J_{1}}\right)\boldsymbol{w}\boldsymbol{+}{\tau}_{RD}\mathbf{g}_{R}^{T}\boldsymbol{\mathrm{\Lambda}}\mathrm{(}{\boldsymbol{P}}_{R})\mathbf{g}_{R}^{*}+{\sigma}^{2}}\right),\right.
\end{gather}
with ${\boldsymbol{\Psi}}_{k}\left(P_{s},P_{J_{1}}\right)\triangleq P_{J_{1}}{k_{J_{1}}^{t}}^{2}{\boldsymbol{\mathrm{\Phi}}}_{Gg}+P_{J_{1}}{k_{R}^{r}}^{2}{\boldsymbol{\mathrm{\Phi}}}_{GG}+P_{s}{k_{s}^{t}}^{2}{\boldsymbol{\mathrm{\Phi}}}_{Gf}+P_{s}{k_{R}^{r}}^{2}{\boldsymbol{\mathrm{\Phi}}}_{GF}+\sigma^{\mathrm{2}}{\boldsymbol{\mathrm{\Phi}}}_{G}$,
with ${\boldsymbol{\mathrm{\Phi}}}_{Gf}\triangleq{\boldsymbol{\mathrm{G}}}_{R}{\boldsymbol{\mathrm{f}}}_{R}{\boldsymbol{\mathrm{f}}}_{R}^{H}{\boldsymbol{\mathrm{G}}}_{R}^{H}$,
${\boldsymbol{\mathrm{\Phi}}}_{Gg}\triangleq{\boldsymbol{\mathrm{G}}}_{R}{\boldsymbol{\mathrm{g}}}_{R}{\boldsymbol{\mathrm{g}}}_{R}^{H}{\boldsymbol{\mathrm{G}}}_{R}^{H}$,
${\boldsymbol{\mathrm{\Phi}}}_{GG}\triangleq\mathrm{diag}\left(\left[\left|g_{R_{1}}\right|^{4},\ldots,\left|g_{R_{N}}\right|^{4}\right]\right)$,
${\boldsymbol{\mathrm{\Phi}}}_{GF}\triangleq{\boldsymbol{\mathrm{G}}}_{R}{\boldsymbol{\mathrm{F}}}_{R}{\boldsymbol{\mathrm{F}}}_{R}^{H}{\boldsymbol{\mathrm{G}}}_{R}^{H}$,
${\boldsymbol{\mathrm{\Phi}}}_{G}\triangleq{\boldsymbol{\mathrm{G}}}_{R}{\boldsymbol{\mathrm{G}}}_{R}^{H}$
, and $\mathit{{\tau}_{RD}\boldsymbol{\mathrm{\triangleq}}{k_{R}^{t}}^{2}+{k_{D}^{r}}^{2}}$.

The information leakage at $R_{l}$, i.e., $\mathit{I(y_{R_{l}};x_{s})}$,
is also obtained as:
\begin{equation}
I\left(y_{R_{l}};x_{s}\right)=\frac{1}{2}{{\mathrm{log}}_{\mathrm{2}}\left(1+{\mathrm{\Omega}}_{l}^{D}\right)\ },\label{eq:20-1}
\end{equation}
where $\mathit{{\mathrm{\Omega}}_{l}^{D}\triangleq\frac{P_{s}{\left|{{\mathrm{f}}_{R}}_{l}\right|}^{2}}{P_{J_{1}}{\tau}_{RJ_{1}}{\left|{g_{R}}_{l}\right|}^{2}+P_{s}{\tau}_{RS}{\left|{f_{R}}_{l}\right|}^{2}+{\sigma}^{2}}}$
with $\mathrm{\mathit{\tau_{RS}\triangleq{k_{s}^{t}}^{2}+{k_{R}^{r}}^{2}}}$
and $\mathrm{{\tau}_{\mathit{RJ_{1}}}\triangleq1+{\mathit{k_{J_{1}}^{t}}}^{2}+{\mathit{k_{R}^{r}}}^{2}}$,
stands for the measured SINR at $\mathit{R_{l}}$ in the presence
of jammer node $\mathit{D}$. 

Notably, while each of the untrusted relays adopts the selection combining
(SC) technique to extract the information symbol based on its own
observation \citep{moradikia2018joint}, Eve attempts to get more
information through combining its observations from both phases. Hence,
considering that Eve sees an equivalent $\mathit{1\times\mathit{\textrm{2}}N_{E}}$
SIMO system, the corresponding information leakage $\mathit{I\left(y_{E};x_{s}\right)}$,
is:
\begin{equation}
I\left(y_{E};x_{s}\right)=\frac{1}{2}{{\mathrm{log}}_{\mathrm{2}}\left[{\mathrm{det}\left({\boldsymbol{\mathrm{I}}}_{2N_{E}}+{\boldsymbol{\mathrm{H}}}_{E}{\boldsymbol{\mathrm{H}}}_{E}^{H}{\boldsymbol{Q}}_{E}^{-1}\right)\ }\right]},\label{eq:21-1}
\end{equation}

\section{Proposed Secrecy Scheme}

The security issue can be addressed through maximizing the ASR, yielding
the optimal solution. Therefore, the ASR in the presence of both $\mathit{E}$
and untrusted relays is evaluated by \citep{ouyang2017destination}:
\begin{equation}
R_{s}={\mathrm{max}{\left[I\left(y_{D};x_{s}\right)-{\mathop{\mathrm{max}}_{i\in\left\{ {\boldsymbol{R}},E\right\} }I\left(y_{i};x_{s}\right)\ }\right]}^{+}},\label{eq:18}
\end{equation}
where ${\left[a\right]}^{+}=\ \mathrm{max}(0,\ a)$, and $I(.;.)$
denotes the mutual information.

From the perspective of secrecy capacity, the optimal strategy is
to maximize $R_{s}$, i.e., $R_{s}^{max}\triangleq\max{R_{s}}$, by
searching the optimal ${\boldsymbol{w}}$, $\mathbf{P}\triangleq\left[\ P_{J_{1}},P_{J_{2}},\ \ P_{s}\right]^{T}$.
Along this line, substituting (11)-(13) into (10) yields the objective
function which is neither convex nor concave and thus solving the
resultant maximization problem will be difficult. Some numerical methods,
e.g., the gradient method or the Newton\textquoteright s method, can
be utilized to exhaustively search for the local optimum though we
cannot guarantee the optimality of the so-obtained solution. To facilitate
the joint design over ${\boldsymbol{w}}$, $\mathbf{\mathbf{P}}$,
a sub-optimal criteria will be presented in the following section.
With the aim of maximizing the secrecy rate, we wish to increase $I\left(y_{D};x_{s}\right)$
as much as possible while keeping the information leakage at the curious
nodes as small as possible. Towards this end, beamforming by distributed
relay nodes should be designed such that the information leakage at
$E$ in phase II, will be thoroughly eliminated. These will be fulfilled
by designing ${\boldsymbol{w}}$ such that it falls into the null
space of the equivalent channel of the relay link from $S$ to $E$,
i.e. $\mathbf{C}_{E}\mathbf{F}_{R}{\boldsymbol{w}}^{\ast}=\mathbf{0}$.
Thanks to NSB described above, the information leakage at \emph{E}
was completely eliminated in phase II and thus we can set $P_{J_{2}}=0$.
Therefore, we can shorten the vector of $\mathbf{P}$ to $\bar{\mathbf{P}}\triangleq\left[\ P_{J_{1}},\ \ P_{s}\right]^{T}$.
Now, by substituting $\mathbf{\Lambda}(\mathbf{P}_{R})\triangleq\mathbf{W}^{H}\mathbf{\Upsilon}_{k}\left(\bar{\mathbf{P}}\right)\mathbf{W}$,
where we have $\mathbf{\Upsilon}_{k}\left(\bar{\mathbf{P}}\right)\triangleq P_{s}\left(1+\tau_{RS}\right)\mathbf{F}_{R}\mathbf{F}_{R}^{H}+P_{J_{1}}\tau_{RJ_{1}}\mathbf{G}_{R}\mathbf{G}_{R}^{H}+\sigma^{2}\mathbf{I}_{N}$,
into (11) and after some manipulations (11) and (13) can be reformulated
as:
\begin{equation}
I\left(y_{D};x_{s}\right)=\frac{1}{2}{{\mathrm{log}}_{\mathrm{2}}\left(1+\frac{P_{s}{\boldsymbol{w}}^{H}{\boldsymbol{\mathrm{\Phi}}}_{Gf}\boldsymbol{w}}{{\boldsymbol{w}}^{H}{\widetilde{\boldsymbol{\mathrm{\Psi}}}}_{k}\left(\bar{\mathbf{P}}\right)\boldsymbol{w}\boldsymbol{+}{\sigma}^{2}}\right)\ },\label{eq:26-1}
\end{equation}
\begin{gather}
I\left(y_{E};x_{s}\right)=\frac{1}{2}\log_{2}\left(1+P_{s}\mathbf{f}_{E}^{H}\left(\tau_{J_{1}}P_{J_{1}}\mathbf{q}_{E}\mathbf{q}_{E}^{H}+P_{s}{k_{s}^{t}}^{2}\mathbf{f}_{E}\mathbf{f}_{E}^{H}+\sigma^{2}{\boldsymbol{I}}_{N_{E}}\right)^{-1}\mathbf{f}_{E}\right),
\end{gather}
where, ${\widetilde{\mathbf{\Psi}}}_{k}\left(\bar{\mathbf{P}}\right)\triangleq P_{J_{1}}{k_{J_{1}}^{t}}^{2}\boldsymbol{\mathbf{\Phi}}_{Gg}+P_{J_{1}}k_{1}\boldsymbol{\mathbf{\Phi}}_{GG}+P_{s}k_{2}\boldsymbol{\mathbf{\Phi}}_{GF}+P_{s}{k_{s}^{t}}^{2}\boldsymbol{\mathbf{\Phi}}_{Gf}+\sigma^{2}k_{3}\boldsymbol{\mathbf{\Phi}}_{G}$
with $k_{1}\triangleq\tau_{RD}\tau_{RJ_{1}}+{k_{R}^{r}}^{2}$, $k_{2}\triangleq\tau_{RD}\left(1+\tau_{RS}\right)+{k_{R}^{r}}^{2}$
and $k_{3}\triangleq1+\tau_{RD}$. Note that, since the intended receiver
knows the channel associated with itself to relays and the weighted
coefficients matrix by some channel estimation method, we expect the
backward self-interference term is totally canceled at $D$. However
due to existence of HI, we still witness some terms related to AN
$z_{1}$ as, i.e., $P_{J_{1}}{k_{J_{1}}^{t}}^{2}\boldsymbol{\mathbf{\Phi}}_{Gg}+P_{J_{1}}k_{1}\boldsymbol{\mathbf{\Phi}}_{GG}$,
which hampers the secrecy and cannot be simply eliminated. In the
following section, we will describe our proposed joint optimal power
allocation and cooperative beamforming (OPA-CB) strategy.

\section{Proposed Joint OPA-CB Design}

For simplicity, we first assume that $i{^\circ}\mathrm{\text{\ensuremath{\in}}}{\left\{ \mathbf{R},E\right\} }$
stands for the most curious node at which the highest information
leakage has been occurred. Since the information leakage at $E$ in
phase II has been omitted, the equations for the case of $i{^\circ}=E$
will be analogous to the case of $i{^\circ}=R_{l}$. Therefore, in
the following, we suppose $i{^\circ}=E$, and the related analyses
for the other case $i{^\circ}=R_{l}$ will be discussed wherever is
needed. Before getting to the proposed joint OPA-CB design, we define
$\mathbf{H}\triangleq\mathbf{C}_{E}\mathbf{F}_{R}$, and $\mathbf{H}_{\bot}$,
which is the projection matrix onto the null space of $\mathbf{H}$,
i.e., ${\boldsymbol{w}}=\mathbf{H}_{\bot}{\boldsymbol{v}}$ where
${\boldsymbol{v}}\in\mathbb{C}^{\left(N-N_{E}-1\right)\times1}$ is
an arbitrary vector. Now, by inserting $\boldsymbol{w=\mathbf{H}_{\bot}v}$
into (14) and subsequently substituting them into (10), subjected
to individual and total power constraints, the following optimization
problem is formulated:
\begin{gather}
{\boldsymbol{\mathrm{P}}}_{0}:\mathop{\mathrm{max}}_{\boldsymbol{v},\bar{\mathbf{P}}}\:\frac{1}{2}{\mathrm{log}}_{\mathrm{2}}{\left(\frac{1+\frac{P_{s}{\boldsymbol{v}}^{H}{\mho}_{Gf}\boldsymbol{v}}{{\boldsymbol{v}}^{H}\mathbf{\Gamma}\left(\bar{\mathbf{P}}\right)\boldsymbol{v}\boldsymbol{+}{\sigma}^{2}}}{1+P_{s}\mathbf{f}_{E}^{H}\left(\tau_{J_{1}}P_{J_{1}}\mathbf{q}_{E}\mathbf{q}_{E}^{H}+P_{s}{k_{s}^{t}}^{2}\mathbf{f}_{E}\mathbf{f}_{E}^{H}+\sigma^{2}\mathbf{I}_{N_{E}}\right)^{-1}\mathbf{f}_{E}}\right)\ }\label{eq:27-1}
\end{gather}
s.t.$\textit{ }$

\begin{tabular}{>{\raggedright}p{15.7cm}c}
${{\boldsymbol{1}}_{2}^{T}\boldsymbol{\mathrm{\bar{P}}}\boldsymbol{\mathrm{+}}\boldsymbol{v}}^{H}{\overline{\boldsymbol{\mathrm{\Upsilon}}}}_{k}\left(\mathrm{\bar{\mathbf{P}}}\right)\boldsymbol{v}\le Q_{tot}$,
$\qquad$$\qquad$ & $(16-a)$\tabularnewline
${\boldsymbol{v}}^{H}{\overline{\boldsymbol{\mathrm{\Upsilon}}}}_{k}^{l,l}\left(\mathrm{\bar{\mathbf{P}}}\right)\boldsymbol{v}\le Q_{l}$
, $\forall\,l\boldsymbol{\in}\mathcal{L}$,$\qquad$$\qquad$$\qquad$ & $(16-b)$\tabularnewline
$0<P_{J_{1}}\le{\overline{P}}_{J_{1}}$, $0<P_{s}\le P_{T}$.$\qquad$$\qquad$$\qquad$\vspace{0.3cm}
 & $(16-e)$\tabularnewline
\end{tabular}

where $\mho_{Gf}\triangleq\mathbf{H}_{\bot}^{H}\boldsymbol{\Phi}_{Gf}\mathbf{H}_{\bot}$,
$\mathbf{\Gamma}\left(\bar{\mathbf{P}}\right)\triangleq\mathbf{H}_{\bot}^{H}{\widetilde{\mathbf{\Psi}}}_{k}\left(\bar{\mathbf{P}}\right)\mathbf{H}_{\bot},{\bar{\mathbf{\Upsilon}}}_{k}\left(\bar{\mathbf{P}}\right)\triangleq\mathbf{H}_{\bot}^{H}\mathbf{\Upsilon}_{k}\left(\bar{\mathbf{P}}\right)\mathbf{H}_{\bot}$,
${\bar{\mathbf{\Upsilon}}}_{k}^{l.l}\left(\bar{\mathbf{P}}\right)\triangleq\mathbf{H}_{\bot}^{H}\mathbf{\Upsilon}_{k}^{l.l}$$\left(\bar{\mathbf{P}}\right)$
$\mathbf{H}_{\bot}$, with $\mathbf{\Upsilon}_{k}^{l.l}\left(\mathbf{P}\ \right)\triangleq P_{s}\left(1+\tau_{RS}\right)\mathbf{F}_{R}\mathbf{E}_{l}\mathbf{F}_{R}^{H}+P_{J_{1}}\tau_{RJ_{1}}\mathbf{G}_{R}\mathbf{E}_{l}\mathbf{G}_{R}^{H}+\sigma^{2}$
and $\mathbf{E}_{l}\triangleq\mathrm{diag}\left(\mathbf{e}_{l}\right)$
in which the vector $\mathbf{e}_{l}$ denotes a unit vector whose
$n$-th entry equals to one. The objective function (16) and also
the constraints (16-a) and (16-b) are non-convex. As a consequence,
the joint optimization problem (16) is NP-hard and finding a global
optimum is computationally expensive or even intractable. In this
case, computing a local optima via a low-complexity algorithm is more
meaningful, in practice. Along this line, resorting to the SPCA \citep{10.2307/169728},
the non-convex problem (16) is approximated by a sequence of convex
problems that are much easier to be solved. 

\subsection{The Proposed SPCA-based Solution }

Variables ${\boldsymbol{v}},P_{J_{1}}$, and $P_{s}$ have been coupled
with each other which is an impediment in solving the optimization
problem (16). To cope with this issue, the following variable transformation
is introduced:
\begin{equation}
q_{J_{1}}\triangleq\frac{1}{P_{J_{1}}},\ q_{s}\triangleq\frac{1}{P_{s}},\ \ \ \ \textrm{and}\ \ \mathbf{q}\triangleq\left[q_{J_{1}},\ q_{s}\right]^{T},
\end{equation}

With the notation introduced in (17), we can rewrite the power constraints
in (16-a) and (16-b) as:
\begin{gather}
\sigma^{2}{\boldsymbol{v}}^{H}{\boldsymbol{v}}\frac{\tau_{RJ_{1}}{\boldsymbol{v}}^{H}\mathbf{H}_{\bot}^{H}\mathbf{G}_{R}\mathbf{H}_{R}^{H}\mathbf{H}_{\bot}{\boldsymbol{v}}}{q_{J_{1}}}+\mathbf{1}_{2}^{T}\mathbf{q}+\frac{\left(1+\tau_{RS}\right){\boldsymbol{v}}^{H}\mathbf{H}_{\bot}^{H}\mathbf{F}_{R}\mathbf{F}_{R}^{H}\mathbf{H}_{\bot}{\boldsymbol{v}}}{q_{s}}+\le Q_{tot},
\end{gather}
\begin{align}
\sigma^{2}{\boldsymbol{v}}^{H}{\boldsymbol{v}}\frac{\left(1+\tau_{RS}\right){\boldsymbol{v}}^{H}\mathbf{H}_{\bot}^{H}\mathbf{F}_{R}\mathbf{E}_{l}\mathbf{F}_{R}^{H}\mathbf{H}_{\bot}{\boldsymbol{v}}}{q_{s}}+\frac{\tau_{RJ_{1}}{\boldsymbol{v}}^{H}\mathbf{H}_{\bot}^{H}\mathbf{G}_{R}\mathbf{E}_{l}\mathbf{G}_{R}^{H}\mathbf{H}_{\bot}{\boldsymbol{v}}}{q_{J_{1}}}+ & \le Q_{l},\forall\,l\in\mathbf{\mathcal{L},}
\end{align}
where the term $\mathbf{1}_{2}^{T}\mathbf{q}$, that is the summation
of two convex functions, is strictly convex over $\mathbf{q}$, \citep[Sec. 3.2]{10.5555/993483}.
On the other hand, we know that the quadratic form ${\boldsymbol{z}}^{H}{\mathbf{A}\boldsymbol{z}}$
is convex with respect to the variable ${\boldsymbol{z}}$ if the
matrix $\mathbf{A}$ is positive semidefinite \citep[Sec. 4.2]{10.5555/993483}.
Furthermore, for $g>0$ the quadratic-over-linear function $\frac{{\boldsymbol{z}}^{H}{\mathbf{A}z}\ }{g}$
is jointly convex over the variables $\left(\mathbf{z},g\right)$
\citep[sec 3.2.6]{10.5555/993483}. Consequently, given that $\mathbf{H}_{\bot}^{H}\mathbf{F}_{R}\mathbf{F}_{R}^{H}\mathbf{H}_{\bot}\succcurlyeq\mathbf{0}$,
$\mathbf{H}_{\bot}^{H}\mathbf{G}_{R}\mathbf{G}_{R}^{H}\mathbf{H}_{\bot}\succcurlyeq\mathbf{0}$,
${\boldsymbol{v}}^{H}\mathbf{H}_{\bot}^{H}\mathbf{F}_{R}\mathbf{E}_{l}\mathbf{F}_{R}^{H}\mathbf{H}_{\bot}\succcurlyeq\mathbf{0}$,
and ${\boldsymbol{v}}^{H}\mathbf{H}_{\bot}^{H}\mathbf{G}_{R}\mathbf{E}_{l}\mathbf{G}_{R}^{H}\mathbf{H}_{\bot}\succcurlyeq\mathbf{0}$,
the power constraints (18) and (19) are jointly convex in $\left(\mathbf{q},{\boldsymbol{v}}\right)$.
We remark that, although the constraints (18) and (19) are convex,
the objective function is still non-convex. To tackle the nonconvexity,
we exploit SPCA. SPCA is an iterative algorithm in which at each iteration
the relevant non-convex part is surrogated by a well-suited inner
convex subset that approximates the non-convex feasible solution set.
The accuracy of this approximation is boosted iteration by iteration.
To apply SPCA, the non-convex problem (16) should be first transformed
into a suitable form. Thus, by some variable transformations, the
problem (16) is converted into the following equivalent problem:
\begin{equation}
{\boldsymbol{\mathrm{P}}}_{1}:\mathop{\mathrm{max}}_{\begin{array}{c}
t_{B},t_{E},\omega_{B},\omega_{E},\\
\beta,\mathbf{q},\boldsymbol{v},a_{s},a_{J_{1}}
\end{array}}\:{\mathcal{D}}\left(t_{B},t_{E}\right)=\frac{1}{2}{\mathrm{log}}_{\mathrm{2}}\left(1+t_{B}\right)-\frac{1}{2}t_{E}\label{eq:27}
\end{equation}
s.t.$\textit{ }$

\begin{tabular}{>{\raggedright}p{15.7cm}c}
$t_{E}=\log_{2}{\omega_{E}}$,  & $(20-a)$\tabularnewline
\end{tabular}

\begin{tabular}{>{\raggedright}p{15.7cm}c}
$t_{B}=\frac{\omega_{B}}{a_{s}+a_{J_{1}}+\beta\sigma^{2}+\sigma^{2}},$ & $(20-b)$\tabularnewline
\end{tabular}

\begin{tabular}{>{\raggedright}p{15.7cm}c}
$\omega_{B}=\frac{\boldsymbol{v}^{H}\text{\ensuremath{\mho}}_{Gf}\boldsymbol{v}}{q_{s}}$, & $(20-c)$\tabularnewline
\end{tabular}

\begin{tabular}{>{\raggedright}p{15.7cm}c}
$a_{s}=\frac{{\boldsymbol{v}}^{H}\mathbf{\mathcal{Z}}_{gf}^{k}{\boldsymbol{v}}}{q_{s}},a_{J_{1}}=\frac{k_{1}\boldsymbol{v}^{H}\text{\ensuremath{\mho}}_{GG}\boldsymbol{v}}{q_{J_{1}}}$, & $(20-d)$\tabularnewline
\end{tabular}

\begin{tabular}{>{\raggedright}p{15.7cm}c}
$\beta=k_{3}{\boldsymbol{v}}^{H}\text{\ensuremath{\mho}}_{G}\boldsymbol{v},$ & $(20-e)$\tabularnewline
\end{tabular}

\begin{tabular}{>{\raggedright}p{15.7cm}c}
$\omega_{E}-1=$$\mathit{\frac{\mathbf{f}_{E}^{H}\left(\tau_{J_{1}}\frac{\mathbf{\boldsymbol{q}}_{E}\mathbf{\boldsymbol{q}}_{E}^{H}}{q_{J_{1}}}+{k_{s}^{t}}^{2}\frac{\mathbf{\boldsymbol{f}}_{E}\mathbf{\boldsymbol{f}}_{E}^{H}}{q_{s}}+\sigma^{2}{\boldsymbol{I}}_{N_{E}}\right)^{-1}\mathbf{f}_{E}}{q_{s}}}$\vspace{0.3cm}
 & $(20-f)$\tabularnewline
\end{tabular}

\begin{tabular}{>{\raggedright}p{15.7cm}c}
$\frac{1}{q_{s}}\le P_{T},\:\frac{1}{q_{J_{1}}}\le{\bar{P}}_{J_{1}}$,  & $(20-g)$\tabularnewline
\end{tabular}

\begin{tabular}{>{\raggedright}p{15.7cm}c}
(18), (19), \vspace{0.3cm}
 & $(20-i)$\tabularnewline
\end{tabular} 

where $\mathbf{\mathcal{Z}}_{gf}^{k}\triangleq{k_{s}^{t}}^{2}\mathbf{H}_{\bot}^{H}\mathbf{\Phi}_{Gf}\mathbf{H}_{\bot}+k_{2}\mathbf{H}_{\bot}^{H}\mathbf{\Phi}_{GF}\mathbf{H}_{\bot},\text{\ensuremath{\mho}}_{GG}\triangleq\mathbf{H}_{\bot}^{H}\boldsymbol{\Phi}_{GG}\mathbf{H}_{\bot},\text{\ensuremath{\mho}}_{G}\triangleq\mathbf{H}_{\bot}^{H}\boldsymbol{\Phi}_{G}\mathbf{H}_{\bot}$. 

Since the objective function of ${\boldsymbol{\mathrm{P}}}_{1}$ is
a combination of concave function ${\mathrm{log}}_{\mathrm{2}}\left(1+t_{B}\right)$
and a linear function $t_{E}$, we deal with a concave objective function.
However, the equality constraints (20-a)-(20-f) are still non-convex
because of having the functions on both sides of the equalities which
are not affine. As such, these non-convex equality constraints should
be first transformed into the equivalent convex inequality constraints
to make it more tractable. Subsequently, the problem (20) becomes:
\begin{equation}
{\boldsymbol{\mathrm{P}}}_{2}:\mathop{\mathrm{max}}_{\begin{array}{c}
t_{B},t_{E},\omega_{B},\omega_{E},\\
\beta,\mathbf{q},\boldsymbol{v},a_{s},a_{J_{1}}
\end{array}}\:{\mathcal{D}}\left(t_{B},t_{E}\right)\label{eq:27-2}
\end{equation}
s.t.$\textit{ }$

\begin{tabular}{>{\raggedright}p{15.7cm}c}
$t_{E}\geq\log_{2}{\omega_{E}}$,  & $(21-a)$\tabularnewline
\end{tabular}

\begin{tabular}{>{\raggedright}p{15.7cm}c}
$t_{B}\leq\frac{\omega_{B}}{a_{s}+a_{J_{1}}+\beta\sigma^{2}+\sigma^{2}},$ & $(21-b)$\tabularnewline
\end{tabular}

\begin{tabular}{>{\raggedright}p{15.7cm}c}
$\omega_{B}\leq\frac{\boldsymbol{v}^{H}\text{\ensuremath{\mho}}_{Gf}\boldsymbol{v}}{q_{s}}$, & $(21-c)$\tabularnewline
\end{tabular}

\begin{tabular}{>{\raggedright}p{15.7cm}c}
$a_{s}\geq\frac{{\boldsymbol{v}}^{H}\mathbf{\mathcal{Z}}_{gf}^{k}{\boldsymbol{v}}}{q_{s}},a_{J_{1}}\geq\frac{\boldsymbol{v}^{H}\text{\ensuremath{\mho}}_{GG}\boldsymbol{v}}{q_{J_{1}}}$, & $(21-d)$\tabularnewline
\end{tabular}

\begin{tabular}{>{\raggedright}p{15.7cm}c}
$\beta\geq k_{3}{\boldsymbol{v}}^{H}\text{\ensuremath{\mho}}_{G}\boldsymbol{v},$ & $(21-e)$\tabularnewline
\end{tabular}

\begin{tabular}{>{\raggedright}p{15.7cm}c}
$\omega_{E}-1\geq\frac{\mathbf{f}_{E}^{H}\left(\tau_{J_{1}}\frac{\mathbf{\boldsymbol{q}}_{E}\mathbf{\boldsymbol{q}}_{E}^{H}}{q_{J_{1}}}+{k_{s}^{t}}^{2}\frac{\mathbf{\boldsymbol{f}}_{E}\mathbf{\boldsymbol{f}}_{E}^{H}}{q_{s}}+\sigma^{2}{\boldsymbol{I}}_{N_{E}}\right)^{-1}\mathbf{f}_{E}}{q_{s}}$,\vspace{0.3cm}
 & $(21-f)$\tabularnewline
\end{tabular}

\begin{tabular}{>{\raggedright}p{15.7cm}c}
$\frac{1}{q_{s}}\le P_{T},\:\frac{1}{q_{J_{1}}}\le{\bar{P}}_{J_{1}}$,  & $(21-g)$\tabularnewline
\end{tabular}

\begin{tabular}{>{\raggedright}p{15.7cm}c}
(18), (19), \vspace{0.3cm}
 & $(21-i)$\tabularnewline
\end{tabular}

The proof of equivalence between (20) and (21) is provided in Supplementary
material, Section IV.\footnote{Due to pages limits, the detailed convergence \textcolor{black}{analysis}
of SPCA scheme are given in the supplementary document.} 

So far, we have already transformed the non-convex objective function
of the original problem into a concave one ${\boldsymbol{\mathrm{P}}}_{2}$,
while the difficulties now lie in the non-convex constraints (21-b)-(21-f).
Now, to circumvent the difficulties associated with non-convex constraints
(21-c), we first define slack variables $u_{1}\triangleq\mathfrak{Ne}\left\{ \mathbf{H}_{\bot}^{H}\mathbf{G}_{R}\mathbf{f}_{R}\right\} $,
$u_{2}\triangleq\mathfrak{I}m\left\{ \mathbf{H}_{\bot}^{H}\mathbf{G}_{R}\mathbf{f}_{R}\right\} $,
and $\mathbf{u}\triangleq\left[u_{1},u_{2}\right]^{T}$, by which
it can be reformulated as $\frac{\mathbf{u}^{T}\mathbf{u}}{q_{s}}\geq\omega_{B}$.
This form is still non-convex. Concerning the constraint (21-f), because
of the convex function $\frac{1}{q_{s}}$, it cannot be reformulated
into an equivalent linear matrix inequality (LMI) \citep{10.5555/993483}.
By exploiting Schur complement, it can be shown that (21-f) can be
expressed through the following matrix form:
\begin{equation}
\left[\begin{matrix}\boldsymbol{\varpi} & m_{s}\mathbf{f}_{E}\\
m_{s}\mathbf{f}_{E}^{H} & \omega_{E}-1
\end{matrix}\right]\succcurlyeq0,
\end{equation}
\begin{equation}
m_{J_{1}}\le\frac{1}{q_{J_{1}}},
\end{equation}
\begin{equation}
m_{s}\geq\frac{1}{\sqrt{q_{s}}},
\end{equation}
where $\boldsymbol{\varpi}\triangleq\tau_{J_{1}}m_{J_{1}}\mathbf{q}_{E}\mathbf{q}_{E}^{H}+{k_{s}^{t}}^{2}m_{s}^{2}\mathbf{f}_{E}\mathbf{f}_{E}^{H}+\sigma^{2}\mathbf{I}_{N_{E}}$.
Based on above change of variables and equivalent constraints (22)
and (24), the problem (21) is written into an equivalent form as follows:
\begin{equation}
{\boldsymbol{\mathrm{P}}}_{3}:\mathop{\mathrm{max}}_{\mathbf{x}}\:{\mathcal{D}}\left(t_{B},t_{E}\right)
\end{equation}
s.t.$\textit{ }$

\begin{tabular}{>{\raggedright}p{15.7cm}c}
$t_{E}\geq\log_{2}{\omega_{E}}$, $\qquad$$\quad$ & $(25-a)$\tabularnewline
\end{tabular}

\begin{tabular}{>{\raggedright}p{15.7cm}c}
$\omega_{B}\geq a_{s}t_{B}+a_{J_{1}}t_{B}+t_{B}\beta\sigma^{2}+t_{B}\sigma^{2}$$\qquad$$\quad$ & $(25-b)$\tabularnewline
\end{tabular}

\begin{tabular}{>{\raggedright}p{15.7cm}c}
$\begin{array}{c}
u_{1}\triangleq\mathfrak{Ne}\left\{ \mathbf{H}_{\bot}^{H}\mathbf{G}_{R}\mathbf{f}_{R}\right\} ,\,u_{2}\triangleq\mathfrak{I}m\left\{ \mathbf{H}_{\bot}^{H}\mathbf{G}_{R}\mathbf{f}_{R}\right\} ,\,\mathbf{u}\triangleq\left[u_{1},u_{2}\right]^{\text{T}},\end{array}$$\qquad$$\quad$ & $(25-c)$\tabularnewline
\end{tabular}

\begin{tabular}{>{\raggedright}p{15.7cm}c}
$\frac{\mathbf{u}^{T}\mathbf{u}}{q_{s}}\geq\omega_{B},$$\qquad$$\quad$ & $(25-d)$\tabularnewline
\end{tabular}

\begin{tabular}{>{\raggedright}p{15.7cm}c}
$a_{s}\geq\frac{{\boldsymbol{v}}^{H}\mathbf{\mathcal{Z}}_{gf}^{k}{\boldsymbol{v}}}{q_{s}},a_{J_{1}}\geq\frac{\boldsymbol{v}^{H}\text{\ensuremath{\mho}}_{GG}\boldsymbol{v}}{q_{J_{1}}}$,$\qquad$$\quad$ & $(25-e)$\tabularnewline
\end{tabular}

\begin{tabular}{>{\raggedright}p{15.7cm}c}
$\beta\geq k_{3}{\boldsymbol{v}}^{H}\text{\ensuremath{\mho}}_{G}\boldsymbol{v},$$\qquad$$\quad$ & $(25-f)$\tabularnewline
\end{tabular}

\begin{tabular}{>{\raggedright}p{15.7cm}c}
$\frac{1}{q_{s}}\le P_{T},\:\frac{1}{q_{J_{1}}}\le{\bar{P}}_{J_{1}}$,
$\qquad$$\quad$ & $(25-g)$\tabularnewline
\end{tabular}

\begin{tabular}{>{\raggedright}p{15.7cm}c}
(18), (19), (22), (23), (24)$\qquad$$\quad$\vspace{0.3cm}
 & $(25-h)$\tabularnewline
\end{tabular} 

where $\mathbf{x}\triangleq\left[t_{B},t_{E},\omega_{B},\omega_{E},\mathbf{u},\beta,\mathbf{q},\boldsymbol{v},a_{s},a_{J_{1}},m_{s},m_{J_{1}}\right]^{\mathrm{T}}.$
Because of the nonconvexity of the constraints (25-a)-(25-f) and (23),
the problem (25) is still non-convex. As such, to prepare (25) for
using SPCA, we first 
\begin{algorithm}[tbh]
\caption{$\textbf{Joint OPA-CB design algorithm:}$}
$\textbf{Input: }$Set the threshold value for accuracy $\left({\delta}_{I}\right)$
and the maximum number of iterations $\left(N_{max}\right)$

$\textbf{Initialization: }$Initialize ${\mathbf{x}}^{(0)}$. Set
the iteration number $i=0$

$\textbf{Calculating the optimal: }$$\,P_{s}^{^{\circ}}$, $P_{J_{1}}^{^{\circ}}$,${\boldsymbol{v}}^{^{\circ}}$

$\textbf{While }$$\left\{ \left|{\mathcal{D}}\left(t_{B}^{\boldsymbol{(}i+1\boldsymbol{)}},t_{E}^{\boldsymbol{(}i+1\boldsymbol{)}}\right)-{\mathcal{D}}\left(t_{B}^{\boldsymbol{(}i\boldsymbol{)}},t_{E}^{\boldsymbol{(}i\boldsymbol{)}}\right)\right|\boldsymbol{\le}{\delta}_{I}\mathrm{\ or}\begin{array}{c}
\\
\\
\end{array}\right.$

$\quad\quad$$\left.\quad\quad\quad\mathrm{\begin{array}{c}
\\
\\
\end{array}\:}i\le N_{max}\right\} $

$\textbf{ do (1) to (4):}$

$\mathllap{}$$\textbf{ (1).}$Calculate (26)-(29),

$\mathllap{}$$\textbf{ (2).}$ Solve (30), then assign the solution
to $\mathbf{x}^{(i+1)}$,

$\mathllap{}$$\textbf{ (3).}$ Update the slack variables $\gamma\left(i\right),\theta\left(i\right),\rho\left(i\right)$
based on $\mathbf{x}^{(i+1)}$,

$\mathllap{}$$\textbf{ (4).}$ $i=i+1$

$\textbf{ End While,}$

$\textbf{Output:}$ $P_{s}^{^{\circ}},$ $P_{J_{1}}^{^{\circ}}$ ,
${\boldsymbol{v}}^{\boldsymbol{^{\circ}}}$
\end{algorithm}
construct a suitable inner convex subset to approximate the non-convex
feasible solution set. Along this line, if we denote the optimal solutions
of the convex approximation program at the $(i-1)$-th iteration by
$\omega_{E}(i-1)$, $a_{s}\left(i-1\right),a_{J_{1}}\left(i-1\right),t_{B}\left(i-1\right),\beta\left(i-1\right),\mathbf{u}(i-1),\ q_{J_{1}}\left(i-1\right),\textrm{and }q_{s}(i-1)$,
these non-convex constraints (25-a)-(25-f) and (23) can be approximated
by their first-order Taylor approximations around the optimal solutions
at the $(i-1)$-th iteration, and their equivalent are respectively
given in (26)-(29), as follow:
\begin{gather}
\Gamma\left(\omega_{E},\omega_{E}\left(i-1\right)\right)\triangleq\log_{2}{\left(\omega_{E}\left(i-1\right)\right)+\frac{\omega_{E}-\omega_{E}\left(i-1\right)}{\omega_{E}\left(i-1\right).\ln{\left(2\right)}}}\le t_{E},
\end{gather}
\begin{gather}
\digamma(t_{B},a_{s},a_{J_{1}},\beta,\gamma\left(i\right),\theta\left(i\right),\rho\left(i\right),\omega_{B})\triangleq\mathbf{\boldsymbol{\Xi}}\left(t_{B},a_{s},\theta\left(i\right)\right)+\mathbf{\Xi}\left(t_{B},a_{J_{1}},\rho\left(i\right)\right)+\boldsymbol{\Xi}\left(t_{B},\beta,\gamma\left(i\right)\right){\sigma}^{2}+\nonumber \\
{t_{B}\sigma}^{2}-\omega_{B}\le0,
\end{gather}
\begin{gather}
\mathcal{S}\left(\mathbf{u},q_{s};\mathbf{u}\left(i-1\right),q_{s}\left(i-1\right)\right)\triangleq\frac{{\mathbf{u}\left(i-1\right)}^{T}\mathbf{u}\left(i-1\right)}{q_{s}\left(i-1\right)}\times\left(1-\frac{q_{s}-q_{s}\left(i-1\right)}{q_{s}\left(i-1\right)}\ \right)+\nonumber \\
\frac{2{\mathbf{u}\left(i-1\right)}^{T}}{q_{s}\left(i-1\right)}\left(\mathbf{u}-\mathbf{u}\left(i-1\right)\right)\geq\omega_{B},
\end{gather}
\begin{gather}
\mathbf{\mathit{\Upsilon}}\left(q_{J_{1}},q_{J_{1}}\left(i-1\right)\right)\triangleq\frac{1}{q_{J_{1}}\left(i-1\right)}\left(1-\frac{q_{J_{1}}-q_{J_{1}}\left(i-1\right)}{q_{J_{1}}\left(i-1\right)}\ \right)\geq m_{J_{1}},
\end{gather}
where, $\mathbf{\boldsymbol{\mathbf{\Xi}}}\left(x_{1},x_{2},\lambda\right)\triangleq\frac{\lambda}{2}{x_{1}}^{2}+\frac{1}{2\lambda}{x_{2}}^{2},\ \theta\left(i\right)\triangleq\frac{a_{s}\left(i-1\right)}{t_{B}\left(i-1\right)},\rho\left(i\right)\triangleq\frac{a_{J1}\left(i-1\right)}{t_{B}\left(i-1\right)},\gamma\left(i\right)\triangleq\frac{\beta\left(i-1\right)}{t_{B}\left(i-1\right)}$.
Given the above approximations
\begin{algorithm}[tbh]
\textcolor{black}{\caption{$\textbf{The proposed SPCA-based FIPSA:}$}
}

$\textbf{Input: }$Set the threshold value for accuracy $\left({\delta}_{\epsilon}\right)$
and the maximum number of iterations $\left(M_{max}\right)$

$\textbf{Initialization: }$ Initialize with an arbitrary random point
${\mathbf{x}}^{(0)}$ and set the iteration number $i=0$

$\textbf{While }$$ $$\left\{ \left|s^{(i+1)}-s^{(i)}\right|\boldsymbol{\le}{\delta}_{\epsilon}\mathrm{\ or\ }i\le M_{max}\mathrm{\ }\right\} $~$\textbf{ do (1) to (3)}$:

\,$\textbf{ (1).}$ Calculate (26)-(29),

\,$\textbf{ (2).}$ Solve the problem (31),

\,$\textbf{ (3).}$ $i=i+1$

$\textbf{ End While,}$

$\textbf{Output:}$ ${\mathbf{x}}^{\boldsymbol{^{\circ}}}$ ,${s}^{\boldsymbol{^{\circ}}}$.
\end{algorithm}
, the proposed iterative algorithm for the joint OPA-CB design is
presented in Algorithm $\mathrm{I}$, in which the following convex
optimization is solved at the $i$-th iteration: 
\begin{equation}
{\boldsymbol{\mathrm{P}}}_{i}:\mathop{\mathrm{max}}_{\begin{array}{c}
\mathbf{x}\end{array}}\quad{\mathcal{D}}\left(t_{B},t_{E}\right)
\end{equation}
s.t.

\begin{tabular}{>{\raggedright}p{15.7cm}c}
(26)-(29), (18), (19), (22), (24) & $(30-a)$\tabularnewline
$a_{s}\geq\frac{{\boldsymbol{v}}^{H}\mathbf{\mathcal{Z}}_{gf}^{k}{\boldsymbol{v}}}{q_{s}},a_{J_{1}}\geq\frac{\boldsymbol{v}^{H}\text{\ensuremath{\mho}}_{GG}\boldsymbol{v}}{q_{J_{1}}}$,$\qquad\qquad$$\:\:\qquad$ & $(30-b)$\tabularnewline
$\beta\geq k_{3}{\boldsymbol{v}}^{H}\text{\ensuremath{\mho}}_{G}\boldsymbol{v},$ & $(30-c)$\tabularnewline
$\frac{1}{q_{s}}\le P_{T},\:\frac{1}{q_{J_{1}}}\le{\bar{P}}_{J_{1}}$,\footnote{According to \citep{Beck2010} for a constant value $c$, constraints
$\frac{1}{x}\leq c$ and $\frac{1}{x}\geq c$ are convex and concave
functions, respectively. Therefore, constraint $\mathit{\mathrm{(30-d)}}$
is convex and do not need to write Taylor approximation for it.} \vspace{0.2cm}
 & $(30-d)$\tabularnewline
\end{tabular} 

Note that, the iterative process will proceed untill some stopping
criteria is satisfied or the maximum predefined number of iterations
$N_{max}$ is reached. The convergence of the algorithm is investigated
in Supplementary material. 

\subsection{SPCA-based Feasible Initial Points Search Algorithm (FIPSA)}

If the feasible initial points exist for the problem (30), the points
acquired by (30) at each iteration, definitely fall into the feasible
set introduced by the original problem (25) (see Supplementary material,
Lemma 2). However, the feasible initial point may not be found easily,
in general and thus the algorithm may fail at the first iteration
owing to infeasibility. Hence, developing an algorithm to find a feasible
initial point is required. To do so, another optimization problem
is solved in which the real-valued slack parameter $s\geq0$ is minimized.
This parameter can be interpreted as the infeasibility indicator thereby
the violation of constraints of problem (30-a)-(30-b) is measured.
The feasibility problem is given by:
\begin{equation}
{\boldsymbol{\mathrm{P}}}_{5}:\mathop{\mathrm{min}}_{\mathbf{x}}s
\end{equation}
s.t.$\textit{ }$

\begin{tabular}{>{\raggedright}p{15.7cm}c}
$\Gamma\left(\omega_{E},\omega_{E}\left(i-1\right)\right)\leq-s,$$\quad$ & $(31-a)$\tabularnewline
$\digamma(t_{B},a_{s},a_{J_{1}},\beta,\gamma\left(i\right),\theta\left(i\right),\rho\left(i\right),\omega_{B})\leq-s,$ & $(31-b)$\tabularnewline
$\mathcal{S}\left(\mathbf{u},q_{s};\mathbf{u}\left(i-1\right),q_{s}\left(i-1\right)\right)\leq-s,$$\quad$ & $(31-c)$\tabularnewline
$\mathbf{\mathit{\mathit{\Upsilon}}}\left(q_{J_{1}},q_{J_{1}}\left(i-1\right)\right)\leq-s,$$\quad$ & $(31-d)$\tabularnewline
$\frac{1}{\sqrt{q_{s}}}-m_{s}\le-s,$$\quad$ & $(31-e)$\tabularnewline
$\left[\begin{matrix}\boldsymbol{\varpi} & m_{s}\mathbf{f}_{E}\\
m_{s}\mathbf{f}_{E}^{H} & \omega_{E}-1
\end{matrix}\right]\succcurlyeq s,$\vspace{0.2cm}
 & $(31-f)$\tabularnewline
$\frac{{\boldsymbol{v}}^{H}\mathbf{\mathcal{Z}}_{gf}^{k}{\boldsymbol{v}}}{q_{s}}-a_{s}\le-s,\frac{\boldsymbol{v}^{H}\text{\ensuremath{\mho}}_{GG}\boldsymbol{v}}{q_{J_{1}}}-a_{J_{1}}\le-s$, & $(31-g)$\tabularnewline
$k_{3}{\boldsymbol{v}}^{H}\text{\ensuremath{\mho}}_{G}\boldsymbol{v}-\beta\le-s,$$\quad$ & $(31-h)$\tabularnewline
$\frac{1}{q_{s}}-P_{T}\le-s,\:\frac{1}{q_{J_{1}}}-{\bar{P}}_{J_{1}}\le-s$,
$\quad$ & $(31-i)$\tabularnewline
$s\geq0$, $\quad$\vspace{0.2cm}
 & $(31-j)$\tabularnewline
\end{tabular}

The optimal solution of the problem (31) at the $(l-1)$-th iteration
is a feasible solution of the problem (31) at the $l$-th iteration.
Therefore, the optimal value of the objective function in the problem
(31) is non-increasing as the iteration number $l$ increases\textcolor{black}{}
\begin{figure}[tbh]
\centering{}\textcolor{black}{}%
\begin{minipage}[c]{0.47\columnwidth}%
\begin{center}
\textcolor{black}{\includegraphics[viewport=0bp 0bp 466bp 563bp,scale=0.62]{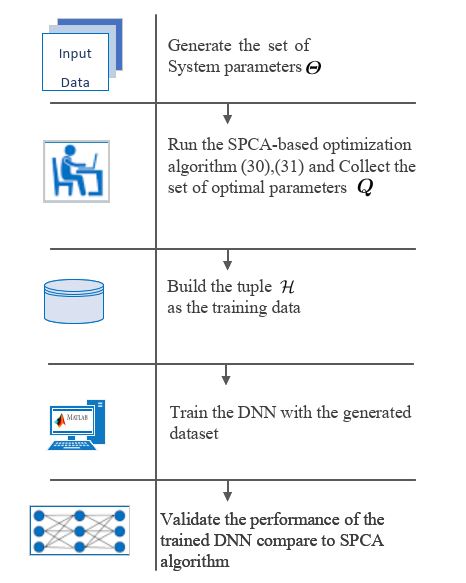}\caption{The process of the proposed DNN-based scheme.}
}
\par\end{center}%
\end{minipage}\textcolor{black}{~~~~~~~}%
\begin{minipage}[c]{0.47\columnwidth}%
\begin{center}
\textcolor{black}{\includegraphics[viewport=0bp 0bp 401bp 581bp,scale=0.6]{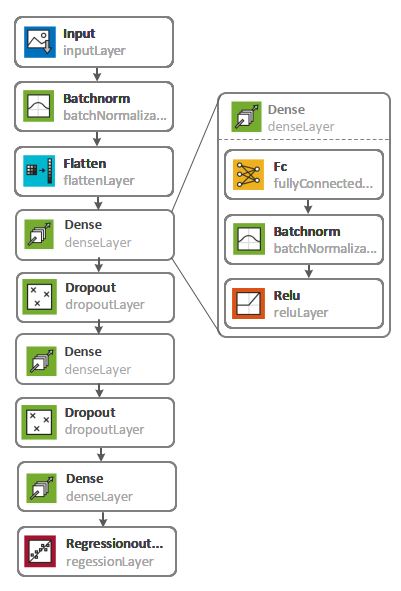}\caption{Proposed DNN framework}
}
\par\end{center}%
\end{minipage}
\end{figure}
 $\left(\mathrm{\textrm{see}}\:\textrm{Lemma}\:\textrm{2}-\boldsymbol{(ii)}\right)$.
Algorithm $2$ is guaranteed to converge.\textcolor{black}{{} }

\textcolor{black}{The SPCA-based feasible initial point search algorithm
is presented in Algorithm $\textrm{2}$. Unlike Algorithm $\textrm{1}$,
where we assumed that it is initialized with a feasible point, Algorithm
$\textrm{2}$ begins with an arbitrary random point. The algorithm
will proceed unless the difference of objective value $\mathit{s}$
in two consecutive iterations becomes smaller than the predefined
threshold value, i.e., $\left|{\mathcal{D}}\left(t_{B}^{\boldsymbol{(}i+1\boldsymbol{)}},t_{E}^{\boldsymbol{(}i+1\boldsymbol{)}},t_{B},t_{E}\right)-{\mathcal{D}}\left(t_{B}^{\boldsymbol{(}i\boldsymbol{)}},t_{E}^{\boldsymbol{(}i\boldsymbol{)}},t_{B},t_{E}\right)\right|\boldsymbol{\le}{\delta}_{\epsilon}$
or the maximum number of affordable iterations is reached. Besides,
whenever the objective value becomes zero the algorithm ceases. Hence,
after calculating the feasible initial points through Algorithm $\textrm{2}$,
the optimal values of $P_{s}^{^{\circ}}$, $P_{J_{1}}^{^{\circ}}$,
${\boldsymbol{v}}^{^{\circ}}$ are obtained via Algorithm $\textrm{\textrm{1}}$.
If no feasible point is obtained for some system parameters, they
should be relaxed so that a feasible solution is achieved. }

\section{\textcolor{black}{Proposed Deep Learning Scheme}}

\textcolor{black}{Facing with high computational load in large-scale
scenarios, the computational complexity order of numerical SPCA-based
solution (30) is significantly increased upon increasing the network
dimension including $N$ and $N_{E}$. Unlike the complex iterative
process of the SPCA-based scheme, in the DNN-based scheme, a multi-layer
model is replaced, where each layer includes some simple matrix multiplication
}
\begin{figure}[tbh]
\centering{}\textcolor{black}{\includegraphics[viewport=0bp 0bp 1079bp 500bp,scale=0.55]{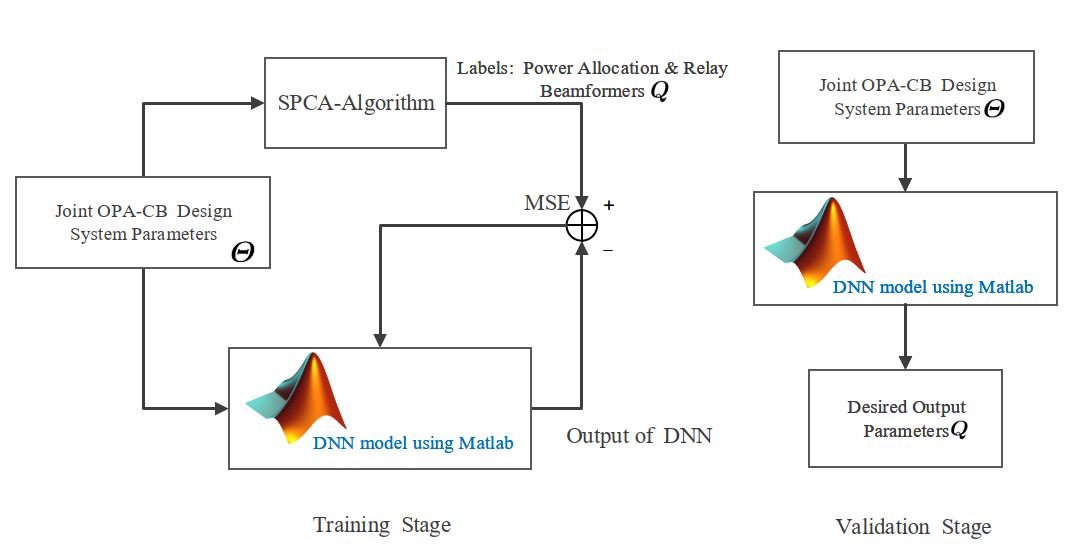}\caption{The training and validation stages of DNN model}
}
\end{figure}
\textcolor{black}{and summation operations followed by a non-linear
mapping, i.e., activation function. This structure guarantees the
real-time performance of the network so that the computational efficiency
of the DNN scheme significantly exceeds the SPCA scheme. The overall
process of the proposed scheme is shown in Fig. 2. we present each
step of this process in detail.}

\subsection{DNN Structure }

The proposed DNN structure is based on multi-layer perceptrons made
up of an input layer, multiple hidden layers, and one output layer.\textcolor{black}{{}
The system parameters $\boldsymbol{\varTheta}=\left[\mathrm{\mathrm{Vec}}\left({\boldsymbol{\mathrm{C}}}_{E}\right)\right.$,
${\mathbf{f}}_{R}$, ${\mathbf{g}}_{R}$, ${\boldsymbol{\boldsymbol{\mathrm{f}}}}_{E}$,
${\boldsymbol{\boldsymbol{\mathrm{q}}}}_{E}$, $N_{E},\,N$, $\left.\boldsymbol{\kappa},Q_{tot,}P_{T,}Q_{l}\right]$
with  $\boldsymbol{\kappa}=\left[k_{s}^{t},k_{J_{1}}^{t},k_{J_{2}}^{t},k_{D}^{r},k_{R}^{r},k_{R}^{t}\right]^{\mathrm{T}},$
is fed to the proposed DNN, }and \textcolor{black}{$\boldsymbol{Q}=\mathrm{\mathit{\left[R_{s},P_{s},P_{J_{1}},\mathbf{{\boldsymbol{w}}^{\mathit{}}}\right]}}$,
}is optained at the output. Each neuron receives information from
the neurons of the preceding layer according following formula:
\begin{gather}
p_{j}^{(i+1)}=\xi^{(i)}\left(\stackrel[k=1]{k=N_{i}}{\sum}w_{j,k}^{(i)}u_{k}^{(i)}+b_{k}^{(i)}\right),
\end{gather}
where $w_{j,k}^{(i)}$ is a weight that determines relationship between
the $k$-th neuron in $i$-th layer and $j$-th neuron in the $(i+1)$-th
layer. Moreover, $b_{k}^{(i)}$ is the bias of the neuron associated
with the $k$-th neuron in the $(i+1)$-th layer. In the $i$-th layer,
the number of neurons is represented by $N_{i}$, and $\xi^{(i)}(.)$
denotes the activation function. \textcolor{black}{Note that the rectified
linear unit (ReLU) is a well-suited activation function for such nonlinear
regression, and therefore is used in the proposed network.} The ReLU
function can mitigate the gradient dispersion, which is denoted as
$\textrm{ReLU}(x)=\textrm{max}(0,x)$.\textcolor{black}{{} }

\subsection{Data Generation Stage}

The proposed DNN is trained in an epochal setting, and the input data
is generated according to the following procedure. First, the system
parameters $\boldsymbol{\varTheta}^{\mathit{(i)}}$ are generated,
where $i$ represents the $i$-th training sample. Then, the optimized
power allocations and relay beamformer $\boldsymbol{\boldsymbol{Q}}^{\mathit{(i)}}$
are generated using SPCA for each tuple $\boldsymbol{\varTheta}^{\mathit{(i)}}$\textcolor{black}{{}
according to (30) and (31)}. The $i$-th training sample is the tuple
$\mathcal{H=\left\{ \left(\boldsymbol{\varTheta}^{\mathit{(i)}}\mathit{,\:}\boldsymbol{\boldsymbol{Q}}^{\mathit{(i)}}\right)\right\} }$.
We then perform this process $K$ times to generate the validation
and training datasets, therefore \textcolor{black}{$K$ is the size
of the dataset}. \textcolor{black}{Before training the neural network,
it is necessary to preprocess the data set with the aim of improving
the network generalization ability and reducing the influence of singular
values \citep{8731532}. Accordingly, normalization is carried out
prior to training the network \citep{9012708}, and thus the data
are normalized through a standard normal distribution given by:
\begin{equation}
\tilde{\boldsymbol{\varTheta}}\triangleq\left\{ \tilde{\boldsymbol{\varTheta}}^{(i)}\right\} _{i=1}^{K},\:\tilde{\boldsymbol{\varTheta}}^{(i)}=\frac{\boldsymbol{\varTheta}^{(i)}-\mu^{(i)}}{\sigma^{(i)}},
\end{equation}
where $\tilde{\boldsymbol{\varTheta}}^{(i)}$, $\mu^{(i)}$, and $\sigma^{(i)}$
express the normalized value, the mean value and the standard deviation
of the $i$-th }training\textcolor{black}{{} sample, respectively. The
proposed DNN framework is shown in Fig. 3.}

\subsection{Training Stage}

The DNN scheme is shown in Fig. 4, which consists of the training
and the validation stages. The training process is responsible for
continuously optimizing the weight of the DNN. \textcolor{black}{The
adaptive Adam-optimizer is also used aimed at alleviating the burden
of debugging parameters such as the learning rate and the batch size.
In addition, a dropout method with probability $p=0.75$ is adopted,
which randomly deactivates some neurons with the aim of elliminating
the dependency on the output of a specific neuron. The overfitting
problem is therefore prevented, resulting in enhancing the model robustness
and improving the scalability of the network.} Moreover, the decay
rate is fixed to $0.9$, and the batch size and the learning rate
are selected using cross-validation\footnote{A dataset can be repeatedly split into a training dataset and a validation
dataset, known as cross-validation. These repeated partitions can
be done in various ways such as dividing into $2$ equal datasets,
and using them as training/validation, and then validation/training,
or repeatedly selecting a random subset as a validation dataset \citep{article}.
To validate the model performance, sometimes an additional test dataset
that was held out from cross-validation is used.}. The goal is to minimize the loss function which reflects the mean
square error (MSE) between the label values and the network output
values. The training and the validation stages are shown in Fig.4.
Moreover, the loss function of the proposed DNN can be rewritten as:
\begin{equation}
\textrm{MSE}=\min_{\left\{ w_{j,k}^{(i)}\right\} ,\left\{ b_{k}^{(i)}\right\} }\sum_{j}\left(\left\Vert p_{j}^{(out)}-\boldsymbol{\boldsymbol{Q}}_{j}^{\mathit{(i)}}\right\Vert ^{2}\right)
\end{equation}
where $p_{j}^{(out)}$ and $\boldsymbol{\boldsymbol{Q}}_{j}^{\mathit{(i)}}$
are the $j$-th entry of output layer and the label value, respectivelly.
In the training stage of DNN, we update the weight $\left\{ w_{j,k}^{(i)}\right\} _{j,k,i}$
and bias $\left\{ b_{k}^{(i)}\right\} _{k,i}$ to minimize loss function.

\subsection{Validation Stage }

The validation process is necessary to tune the hyperparameters (i.e.
the architecture) and provide an unbiased evaluation of the trained
DNN fitted on the training dataset. Clearly, the primary goal of the
proposed DNN is to achieve the best performance on a new dataset.
As a result, the simplest approach to evaluate the performance of
the proposed DNN is to assess the MSE using the data independent of
those used in the training stage. we pass the validation set to the
network and then we get the output of the DNN model. Then, the MSE
of the system is calculated using the label values and the output
values. 

\section{\textcolor{black}{\normalsize{}Complexity Analysis}}

\textcolor{black}{In this subsection, we aim to compare the complexity
order of SPCA-based and the DNN-based schemes, respectively. The overall
proposed SPCA-based scheme involves the Algorithm 1 and the FIPSA
Algorithm 2. As both Algorithm1 and Algorithm 2 (Eq. (30) and (31))
have analogous structures, calculating the complexity order of only
Algorithm 1 is sufficient. The optimization problem (30), used in
Algorithm 1, is a semidefinite programming (SDP), whose all the constraints
were transformed into LMIs by using Schur complements. Even though
it is not a standard SDP problem \citep{10.5555/993483}, using the
interior-point method, the worst-case complexity can be calculated
by $\mathcal{O}\left(m^{2}\left(\sum_{i}m_{i}^{2}\right)\sqrt{\sum_{i}m_{i}}\right)$,
where $m$ is the number of optimization variables and $m_{i}$ is
the dimension of the $i$-th semidefinite cone \citep{doi:10.1080/1055678021000045123}.
Therefore, when the interior-point method is employed to solve the
problem (30), the worst-case computational complexity at each iteration
can be calculated by:
\begin{gather}
\zeta_{\textrm{SPCA}}\triangleq\mathcal{O}\left((N\text{\textminus}N_{E})^{2}\left(2(N\text{\textminus}N_{E}+1)^{2}+(N_{E}+1)^{2}\right)\left.\sqrt{2(N\text{\textminus}N_{E}+1)+(N_{E}+1)}\right),\right.
\end{gather}
}

\textcolor{black}{If we let $T_{1}$ and $T_{2}$ respectively denote
the required numbers of iterations for SPCA and FIPSA algorithms,
the overall complexity order is calculated by $(T_{1}+T_{2})$ times
of $\mathit{\zeta_{\textrm{SPCA}}}$. On the other hand, The proposed
DNN-based scheme relies on supervised learning regression algorithm,comprised
of different number of hidden layers and nodes per layer. The time
complexity of DNN can be represented by floating-point operations
(FLOPs) \citep{8892492}. }
\begin{table}[tbh]
\centering{}\textcolor{black}{\caption{Summary of the proposed DNN structure and Notations}
}%
\begin{tabular}{l|c}
\hline 
\textbf{\textcolor{black}{Symbols}} & \multicolumn{1}{c}{\textbf{\textcolor{black}{Values}}}\tabularnewline[0.2cm]
\hline 
\hline 
\textcolor{black}{Training epoch } & \textcolor{black}{$400$}\tabularnewline
\hline 
\textcolor{black}{Batch size } & \textcolor{black}{$32$}\tabularnewline
\hline 
\textcolor{black}{Learning rate} & \textcolor{black}{$10^{-3}$}\tabularnewline
\hline 
\textcolor{black}{Decay rate} & \textcolor{black}{$0.9$}\tabularnewline
\hline 
\textcolor{black}{The size of training dataset} & \textcolor{black}{$9\times10^{3}$}\tabularnewline
\hline 
\textcolor{black}{The size of validation dataset} & \textcolor{black}{$10^{3}$}\tabularnewline
\hline 
\textcolor{black}{The input dimension} & \textcolor{black}{$\begin{array}{c}
2\times(N+N_{E})+\\
N_{E}\times N+5
\end{array}$}\tabularnewline
\hline 
\textcolor{black}{The output dimension } & \textcolor{black}{$N+3$}\tabularnewline
\hline 
\textcolor{black}{The number of neurons in the $1^{th}$ layer: $N_{1}$} & \textcolor{black}{$256$}\tabularnewline
\hline 
\textcolor{black}{The number of neurons in the $2^{th}$ layer: $N_{2}$} & \textcolor{black}{$256$}\tabularnewline
\hline 
\textcolor{black}{The number of neurons in the $3^{th}$ layer: $N_{3}$} & \textcolor{black}{$128$}\tabularnewline
\hline 
Dataset size: $K$ & $10^{4}$\tabularnewline
\hline 
Dropout: $p$ & $0.75$\tabularnewline
\hline 
\end{tabular}
\end{table}
\textcolor{black}{For each layer of the neural network, the number
of FLOPs can be expressed as:
\begin{equation}
\mathrm{\textrm{FLOPs}}_{i}\triangleq2I_{i}O_{i}
\end{equation}
where $I_{i}$ is the input dimension of the $i$-th layer and $O_{i}$
is the output dimension of the $i$-th layer. Therefore, for our supervised
learning scheme, the number of FLOPs is:
\begin{gather}
\zeta_{\textrm{DNN}}=\mathbin{\sum_{i=1}^{i=3}}\,\mathrm{\textrm{FLOPs}}_{i}=2\left(\left(2\times(N+N_{E})+N_{E}\times N+5\right)N_{1}\left.+N_{1}N_{2}+(3+N_{2}+N)N_{3}\right),\right.
\end{gather}
where $N_{i},\,i=1,2,3$ is the neurons number of $i$-th hidden layer.
Comparing $\mathit{\mathit{\zeta_{\textrm{SPCA}}}}$ and $\mathit{\zeta_{\textrm{DNN}}}$,
it can be explicitly seen that: $\mathit{\zeta_{\textrm{DNN}}\ll\mathit{\zeta_{\textrm{SPCA}}}}$,
as we expected.} 

\section{Simulation Results}

In this section, we assess the proposed schemes using simulations.
Our simulation setting is based on the following adjustment, unless
otherwise stated. The threshold values for the stopping criteria of
Algorithm 1 and of FIPSA are respectively $\delta_{I}=\delta_{\epsilon}={10}^{-3}$
, the impairments at each node are $k_{i}^{t}=k_{i}^{r}=0.08$, the
number of antennas at Eve is $N_{E}=2$, the Gaussian noise power
$\sigma^{2}={10}^{-3},\,Q_{tot}=30\,dB,\,P_{T}=1.5\,Q_{tot},\,N=12$
and $Q_{l}=\frac{2Q_{tot}}{N}$. All simulations were averaged over
$1000$ independent channel realizations. \textcolor{black}{}
\begin{figure}[tbh]
\centering{}\textcolor{black}{}%
\begin{minipage}[t]{0.47\columnwidth}%
\begin{center}
\includegraphics[scale=0.47]{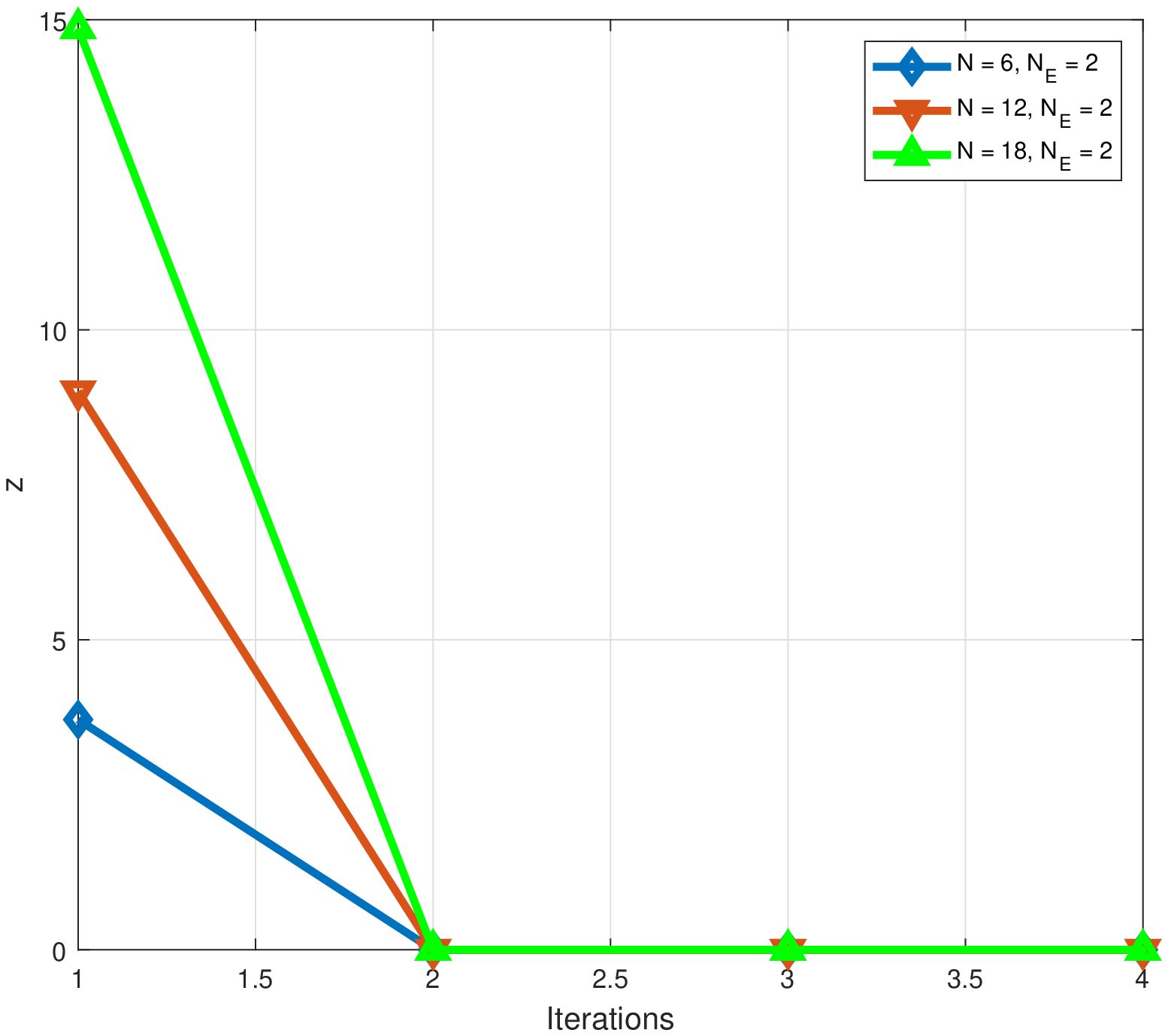}\caption{Evaluating of convergence behavior of the FIPSA through depicting
the OF value in (31) versus the number of iterations for $N=6,12,18$
and $N_{E}=2$}
\par\end{center}%
\end{minipage}\textcolor{black}{~~~~}%
\begin{minipage}[t]{0.47\columnwidth}%
\begin{center}
\includegraphics[scale=0.47]{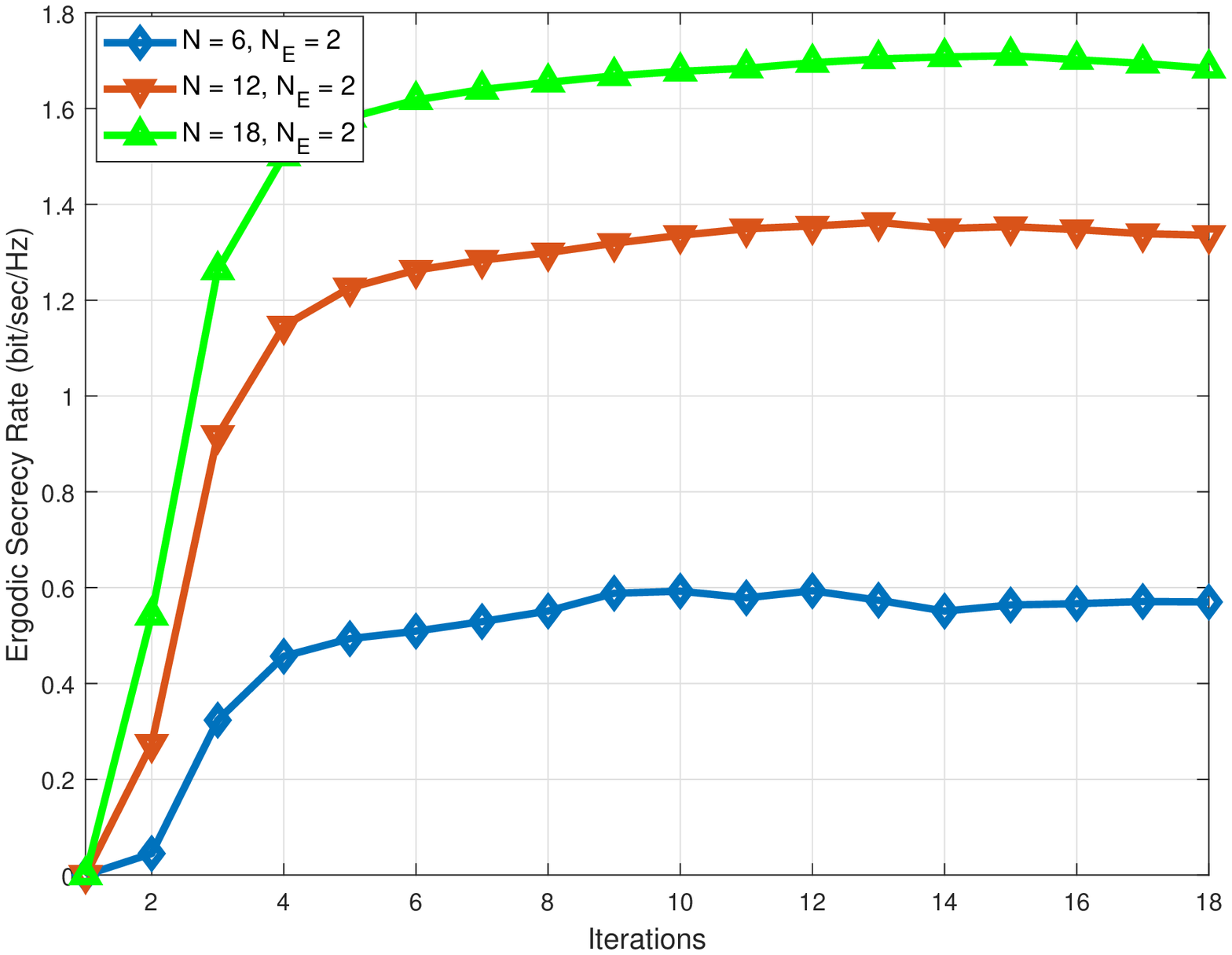}\caption{Average Ergodic secrecy rate $R_{s}$ achieved by the proposed algorithm
versus the number of iterations for $N=6,12,18$ and $N_{E}=2$}
\par\end{center}%
\end{minipage}
\end{figure}
\textcolor{black}{}
\begin{figure}[tbh]
\centering{}\textcolor{black}{}%
\begin{minipage}[t]{0.47\columnwidth}%
\begin{center}
\includegraphics[scale=0.46]{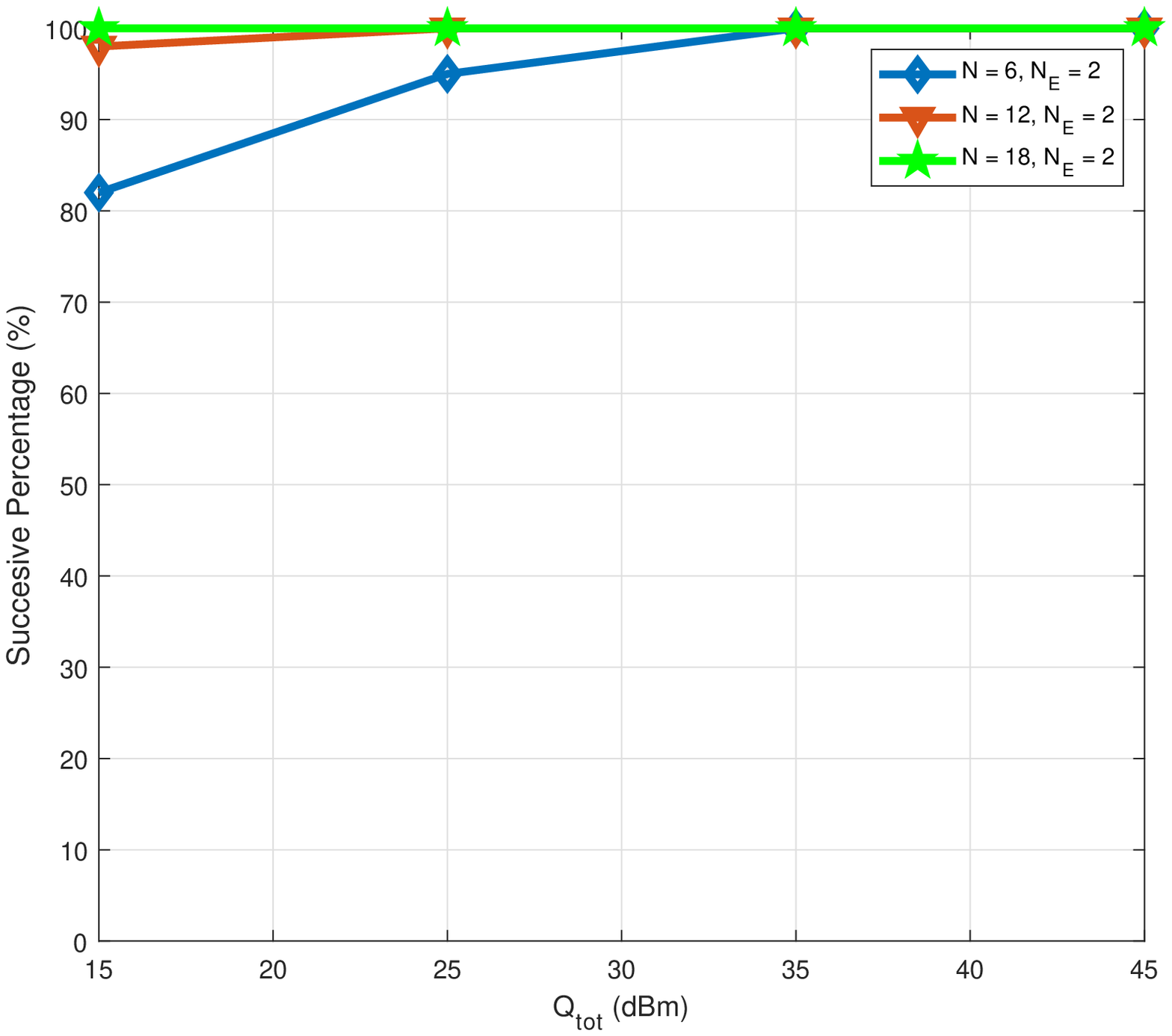}\caption{Percentage of successful cases versus total power budget $Q_{tot}$
for $N=6,12,18$ and $N_{E}=2$}
\par\end{center}%
\end{minipage}\textcolor{black}{~~~~}%
\begin{minipage}[t]{0.47\columnwidth}%
\begin{center}
\includegraphics[scale=0.46]{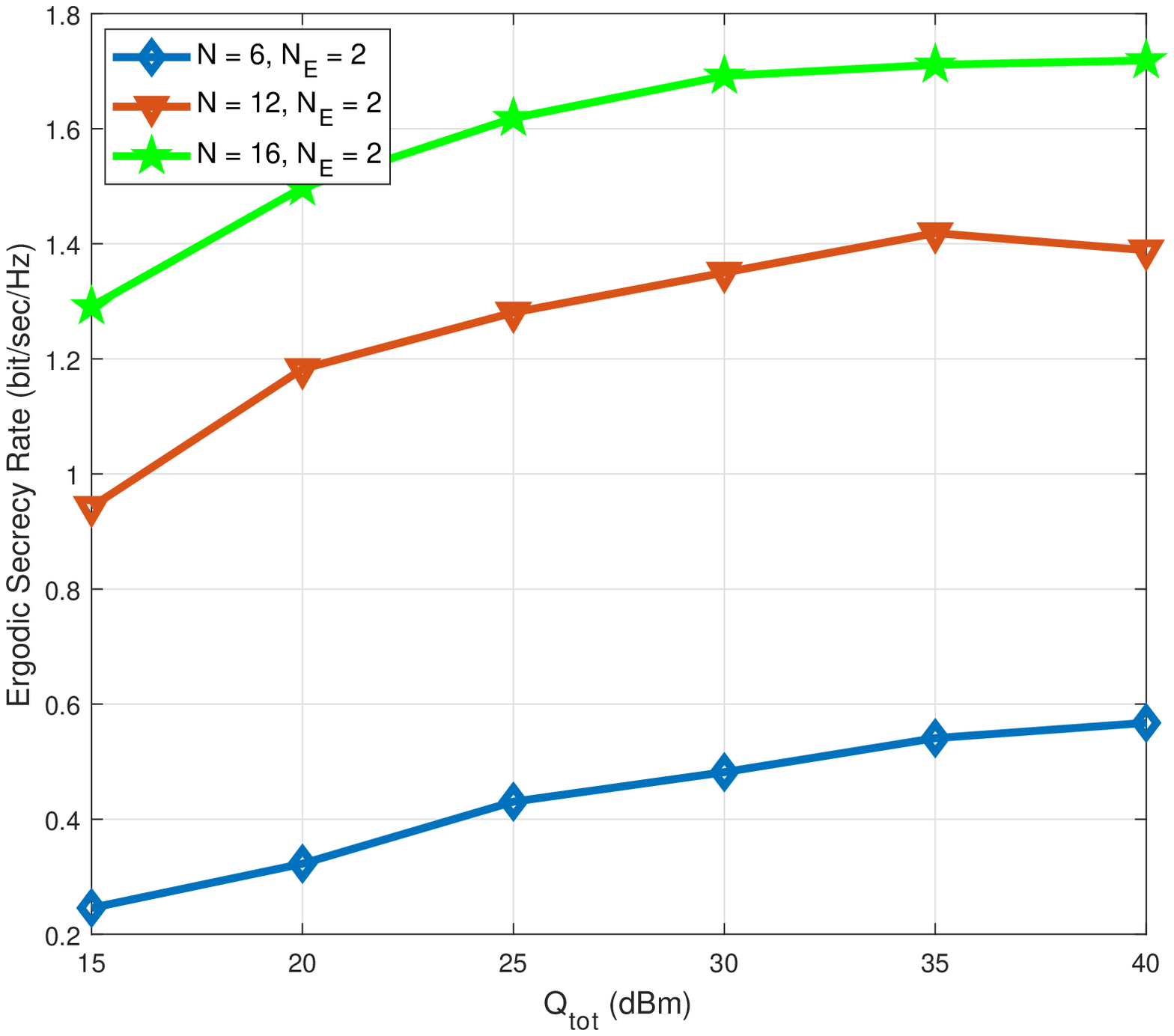}\caption{Average Ergodic secrecy rate $R_{s}$ achived versus the total power
budget $Q_{tot}$ for $N=6,12,18$ and $N_{E}=2$}
\par\end{center}%
\end{minipage}
\end{figure}
\textcolor{black}{The parameters of DNN are set as shown in Table
I. The proposed DNN framework is shown in Fig. 3. This network contains
one input layer, three hidden layers and one output layer, where the
three hidden layers have 256, 256 and 128 neurons. The validation
set is utilized to measure the computing performance and ASR of DNN
and the training set is used for model training. The proposed scheme
is performed in MATLAB $\textrm{2019b}$, with Intel(R) $\textrm{Core(TM)\,i7-7700}$$\mathit{@}$$\textrm{3.6GHz}$,
$\textrm{NVIDIA}$ GeForce $\textrm{GTX}$ 1080.}

Fig. 5 depicts the average convergence of the proposed FIPSA algorithm.
As it can be clearly seen, the average convergence of FIPSA is fast,
such that regardless to the number of relays, it rapidly converges
at the second iteration. The average convergence speed of the proposed
SPCA-based solution for the problem in (25) is shown in Fig. 6. The
results show. the convergence of (31) at almost 10 iterations for
any feasible points. Similar to the convergence behavior of FIPSA,
there is no relationship between the convergence behavior and the
number of relay nodes, which confirms the practicality of our proposed
algorithm. \textcolor{black}{}
\begin{figure}[tbh]
\centering{}\textcolor{black}{}%
\begin{minipage}[c]{0.47\columnwidth}%
\begin{center}
\includegraphics[scale=0.47]{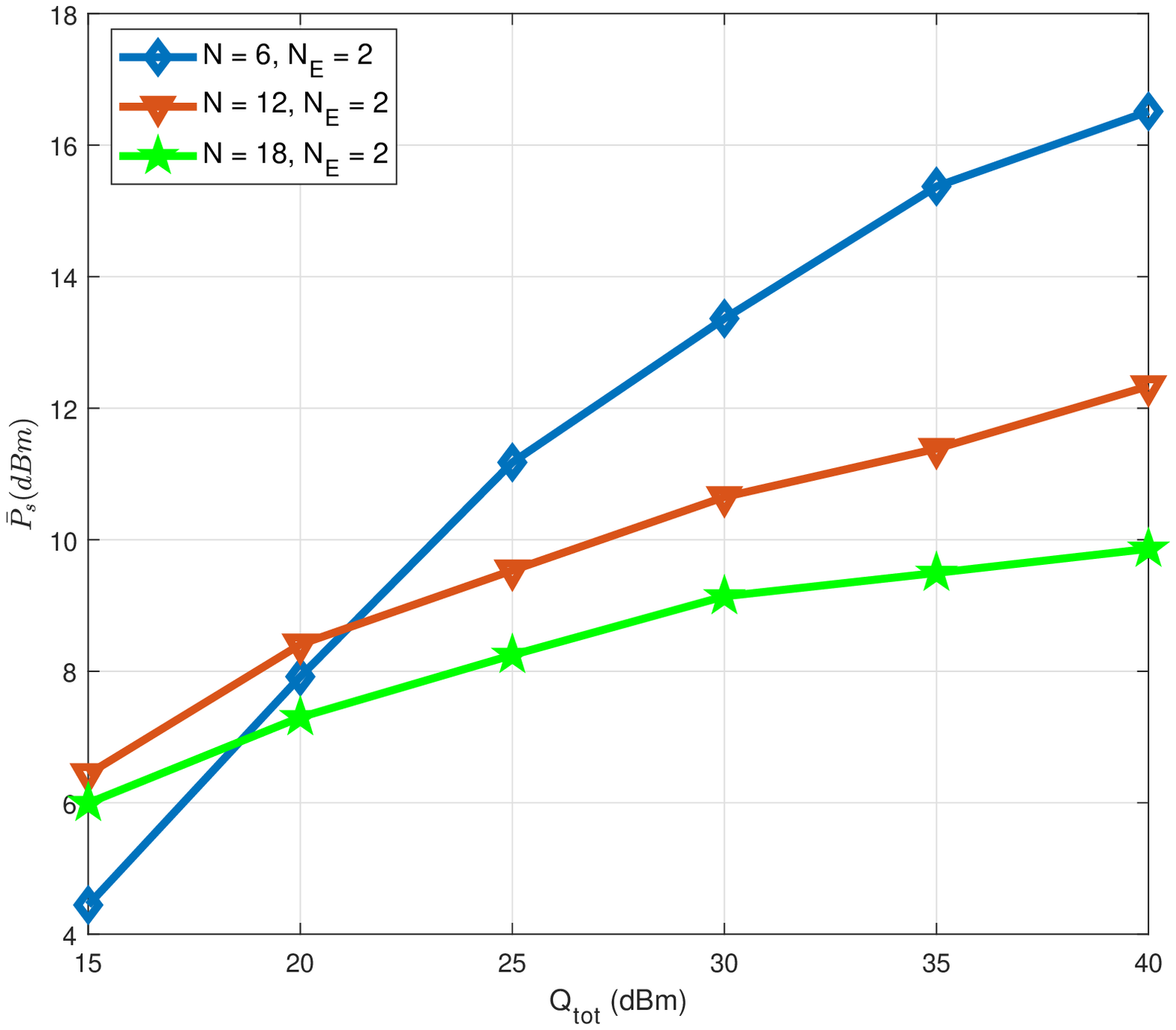}\caption{Average source power consumption $P_{s}$ versus the total power budget
$Q_{tot}$ for $N=6,12,18$ and $N_{E}=2$ }
\par\end{center}%
\end{minipage}\textcolor{black}{~~~~}%
\begin{minipage}[c]{0.47\columnwidth}%
\begin{center}
\includegraphics[scale=0.56]{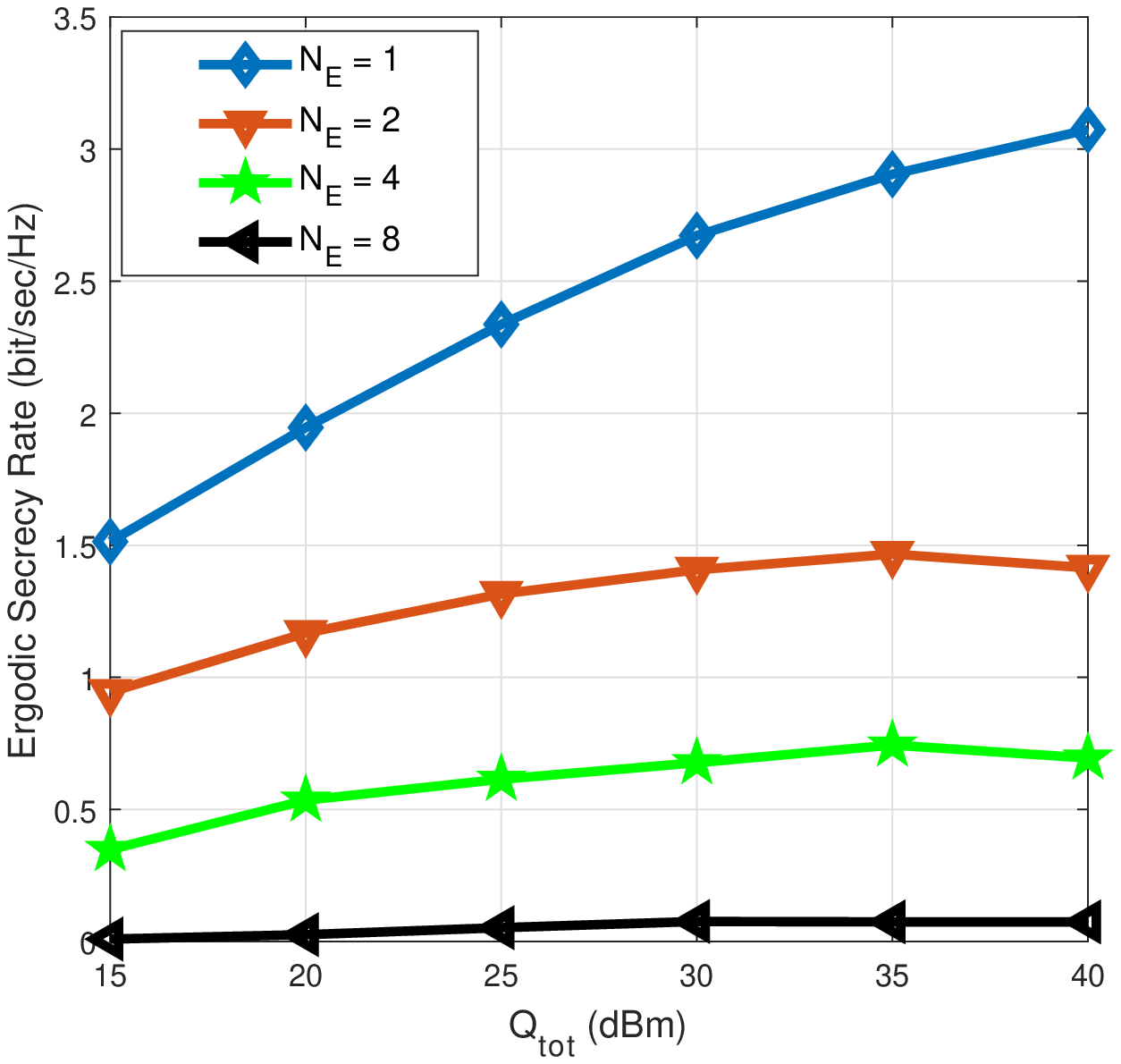}\caption{Ergodic secrecy rate $\mathrm{R_{s}}$ versus the total power budget
$Q_{tot}$ for $N=12$ and $N_{E}=1,2,4,8$ }
\par\end{center}%
\end{minipage}
\end{figure}

In Fig. 7, we depict the percentage of successful cases achieved by
FIPSA versus $Q_{tot}$ for different $\mathit{N}$. Simulation results
show that with increasing $N$ and $Q_{tot}$, the percentage of successful
cases increases. Actually, it can be interpreted that, by increasing
available resources, i.e., transmit power and spatial degree of freedom
(DOF), the feasibility of the non-convex problem (25) would be improved.
Fig. 8 shows the average secrecy rate achieved by SPCA versus the
total power budget for different number of relays. This figure states
that asthe total power becomes high, the average secrecy rate is confined
to a ceiling due to presence of impairments. We can also observe that
given a specific total power budget, the secrecy rate increases as
the number of relays grows. This is because by increasing the number
of relays, the network's degree of freedom is increased, hence enhancing
the ASR.

Observe in Fig. 9, the power consumed by the source for information
transmission is reduced upon increasing the number of untrusted relays.
This observation is originated from two different reasons. On one
hand, by increasing the number of untrusted relays, the information
leakage at phase I is increased. Consequently, to safeguard the information,
most of the total power must be assigned for jamming, and thus the
remaining power for transmitting the information is decreased. On
the other hand, upon increasing $\mathit{N}$,\textcolor{black}{}
\begin{figure}[tbh]
\centering{}\textcolor{black}{}%
\begin{minipage}[t]{0.47\columnwidth}%
\begin{center}
\includegraphics[scale=0.47]{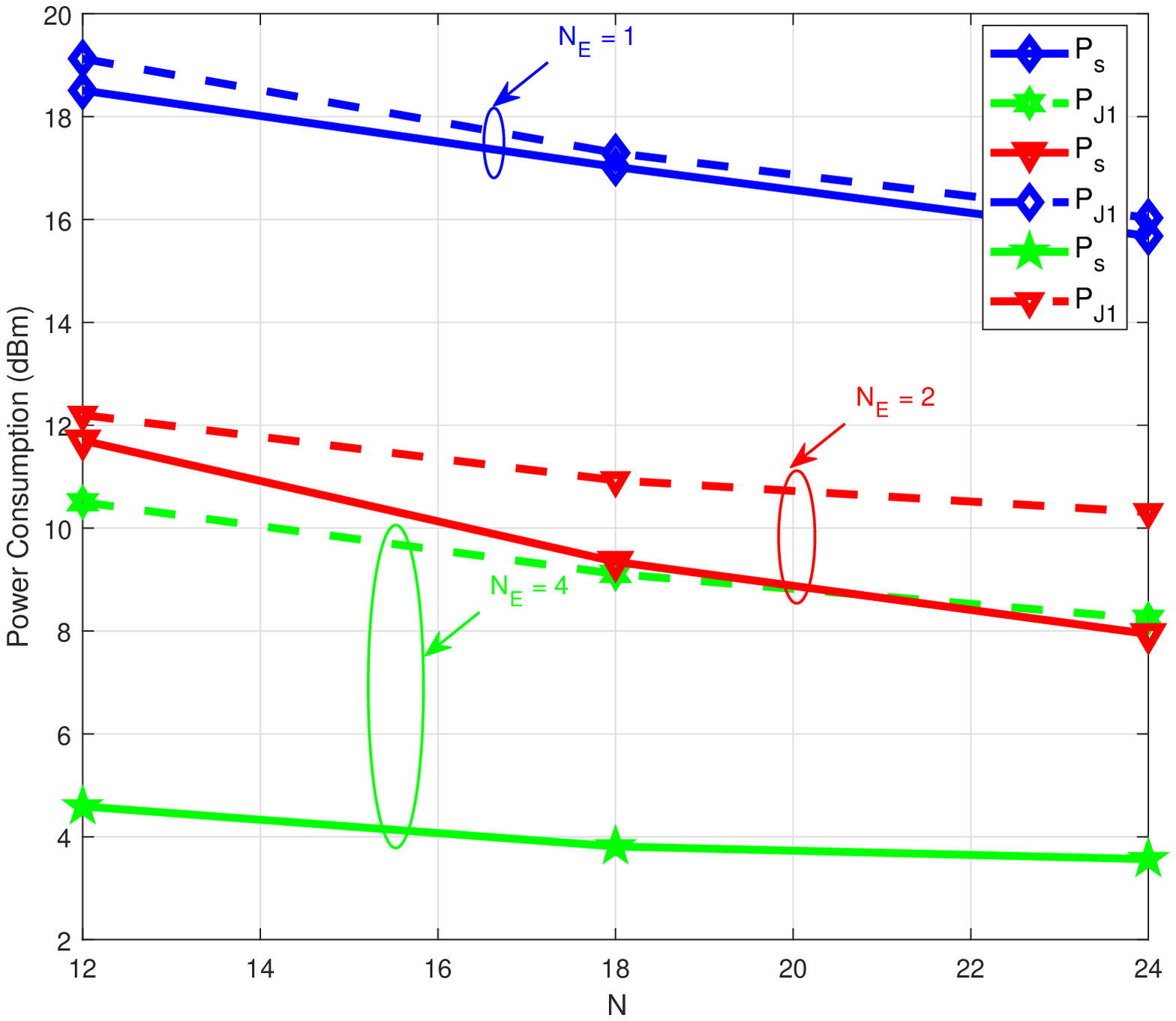}\caption{Average source and friendly jammer power consumption $P_{s},P_{J_{1}}$
versus the total number of relay nodes $N$ for $N_{E}=1,2,4$ }
\par\end{center}%
\end{minipage}\textcolor{black}{~~~~}%
\begin{minipage}[t]{0.47\columnwidth}%
\begin{center}
\includegraphics[scale=0.47]{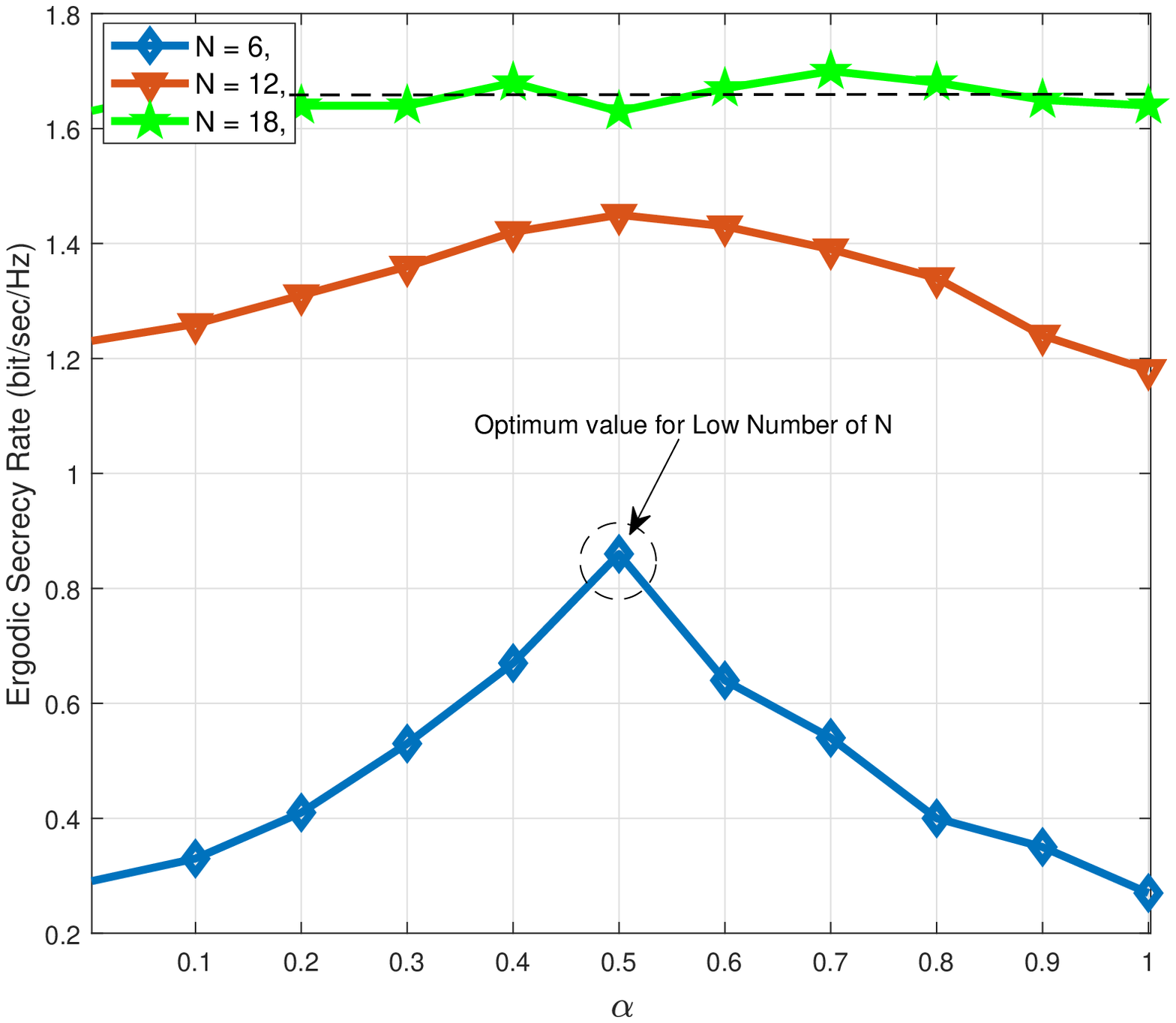}\caption{Average Ergodic secrecy rate $R_{s}$ versus different values of impairments
at relay nodes for $k_{R}^{t}=\alpha$, $k_{R}^{r}=0.2(1-\alpha)$,
$k_{S}^{t}=k_{D}^{r}=0.1$, $N=6,12,18$ and $N_{E}=2$}
\par\end{center}%
\end{minipage}
\end{figure}
\textcolor{black}{{} }more degrees of freedom is provided for the relay
nodes which improves their capability to do beamforming. Hence, there
is no need for increasing $P_{s}$. However at low power budget regimes
most of the power is allocated to relay nodes to perform beamforming.
This will lead to more tangible impacts of HIs at relay nodes. As
a consequence of this intrinsically AN emitted by imperfect relay
nodes, less power is allocated to jamming signal and most of the total
power budget is preferred to be assigned for information transmission. 

The impact of number of the antennas deployed at Eve is shown in Fig.
10. As it would be expected, given a fixed number of relays and a
specific total power budget, the average secrecy rate is decreased
upon increasing $N_{E}$. This is because, deploying more antennas
at the eavesdropper makes it stronger to decipher the information
and hence degrade the secrecy. 

Fig. 11 depicts the power consumed at source and jammer nodes versus
the number of relay nodes. The results were displayed for various
numbers of antennas deployed at Eve. Observe in Fig. 11, the difference
between $P_{s}$ and $P_{J_{1}}$ is increased upon increasing the
number of antennas deployed at Eve. This is because, by increasing
$N_{E}$ the secrecy is more degraded, and hence more jamming power
is required to confront eavesdropping attacks accomplished via external
Eve. Another interesting observation from Fig. 11 is that the impact
of $N_{E}$ on power consumption at source and jammer is much more
considerable than that of the number of cooperative relays. In other
words, the impact of $N_{E}$ is dominant as compared with the number
of untrusted relay nodes. The reason for this is that increasing $N_{E}$
results in purely degrading the security. \textcolor{black}{}
\begin{figure}[tbh]
\centering{}\textcolor{black}{}%
\begin{minipage}[t]{0.47\columnwidth}%
\begin{center}
\includegraphics[scale=0.49]{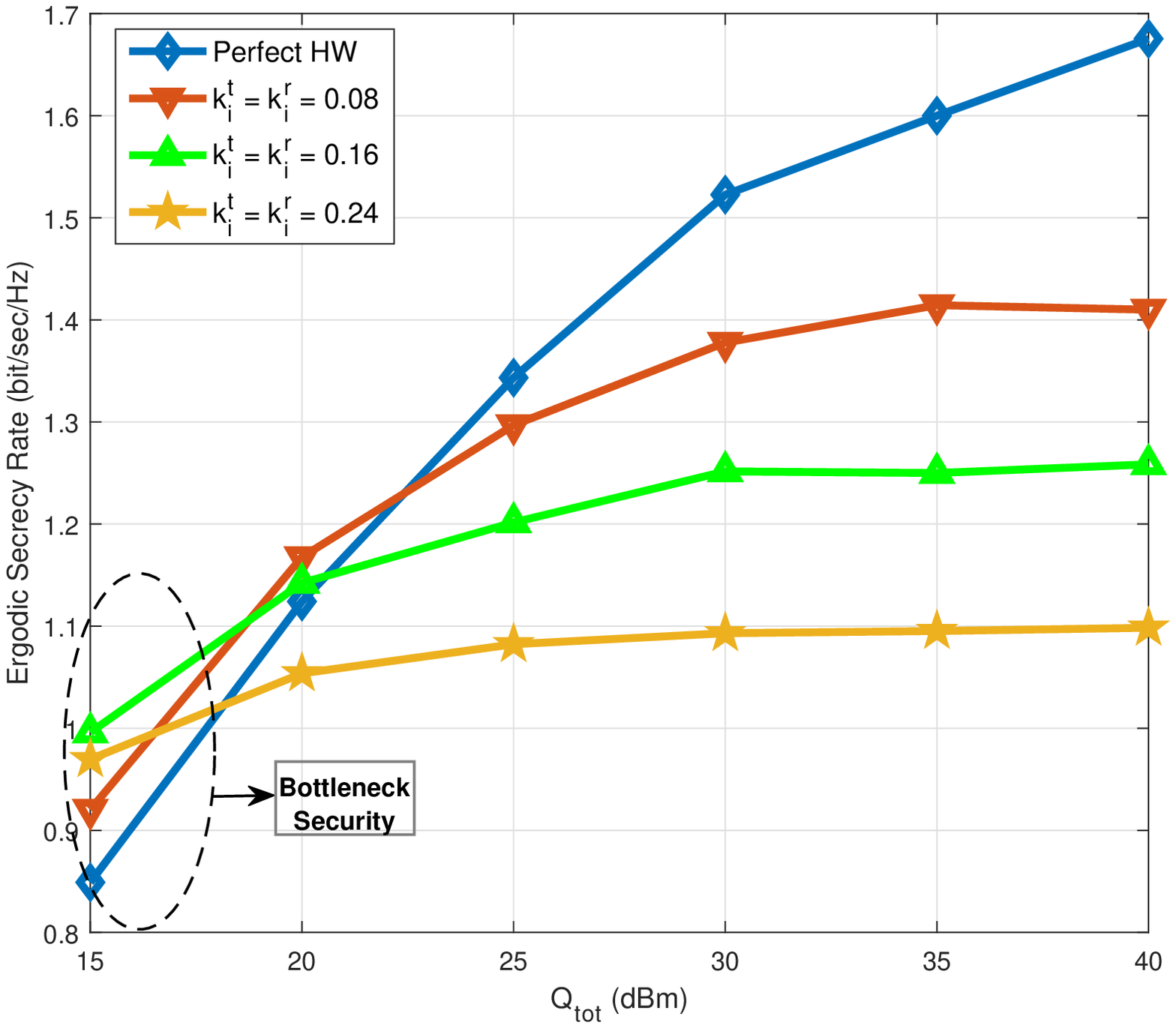}\caption{Average Ergodic secrecy rate $R_{s}$ versus the total power budget
$Q_{tot}$ for different valuses of hardware impairments, $N=12$
and $N_{E}=2$}
\par\end{center}%
\end{minipage}\textcolor{black}{~~~~}%
\begin{minipage}[t]{0.47\columnwidth}%
\begin{center}
\includegraphics[scale=0.62]{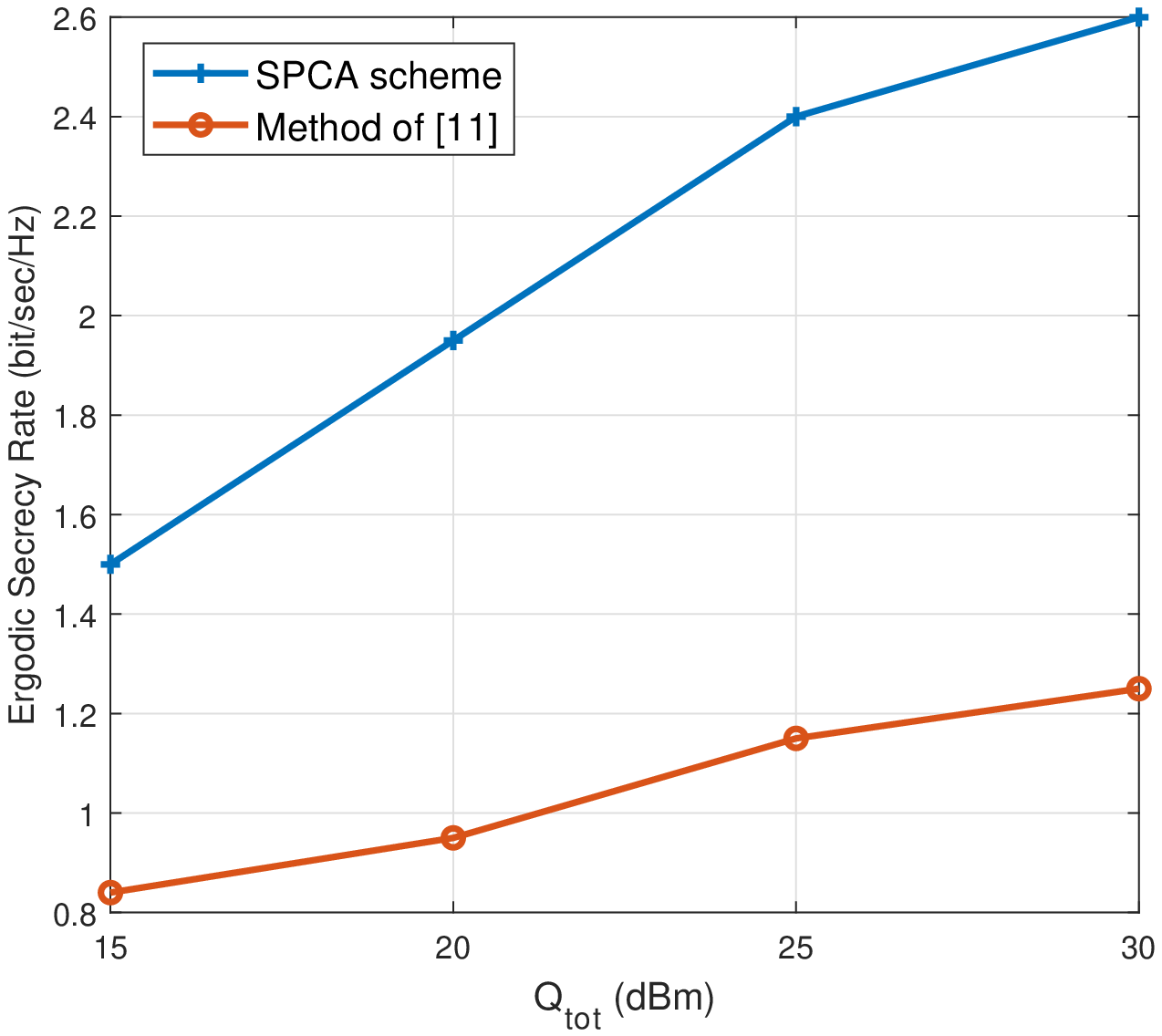}\caption{Average Ergodic secrecy rate $R_{s}$ versus the total power budget
$Q_{tot}$ for minimum quality of service of $\gamma_{min}=12$dB }
\par\end{center}%
\end{minipage}
\end{figure}
However, increasing the number of untrusted relays, although the security
may be jeopardized, the provided DoF together with emitted artificial
noise due to inherent impairment of relay nodes can boost the secrecy
rate, on the other hand. In order to design a practical secure network
we need the engineering perspective of how the total tolerable hardware
impairment at each relay node should be distributed between the RF
reception and RF transmission segments to maximize the secrecy rate.
Actually, depending on the specified expense we aim to spend, it is
needed to know how the RF chain at the transmission and reception
front ends of each relay node should be designed to reach the abovementioned
goal. In this respect, the distribution parameter $0<\alpha<1$ is
defined such that we have $\alpha k_{R}^{r}+\left(1-\alpha\right)k_{R}^{t}=0.2$.
Through this definition, depending the value of $\alpha$, the total
impairment level of 0.2, considered at each relay node, is divided
between the transmission and reception sections of the same node.
Observing Fig. 12, it can be explicitly seen that in the case of using
low number of relays, it is favored to expend our budget at the reception
and transmission front ends equally. This intuitive result has been
mathematically analyzed in \citep{8334689}, as well. By doing so,
both the RF reception and RF transmission experiences equal levels
of impairment. However, upon increasing the number of relays, the
network secrecy performance will be independent of the network HIs.
This is due to the fact that the network's DoF is enhanced upon increasing
the number of relays. 

The impact of impairment levels on the ASR is demonstrated in Fig.
13. As it can be observed, at high power budgets, the more the hardware
are impaired, the less secrecy rate is acquired. In contrast, at low
power budgets, \textcolor{black}{}
\begin{figure}[tbh]
\centering{}\textcolor{black}{}%
\begin{minipage}[t]{0.47\columnwidth}%
\begin{center}
\includegraphics[scale=0.63]{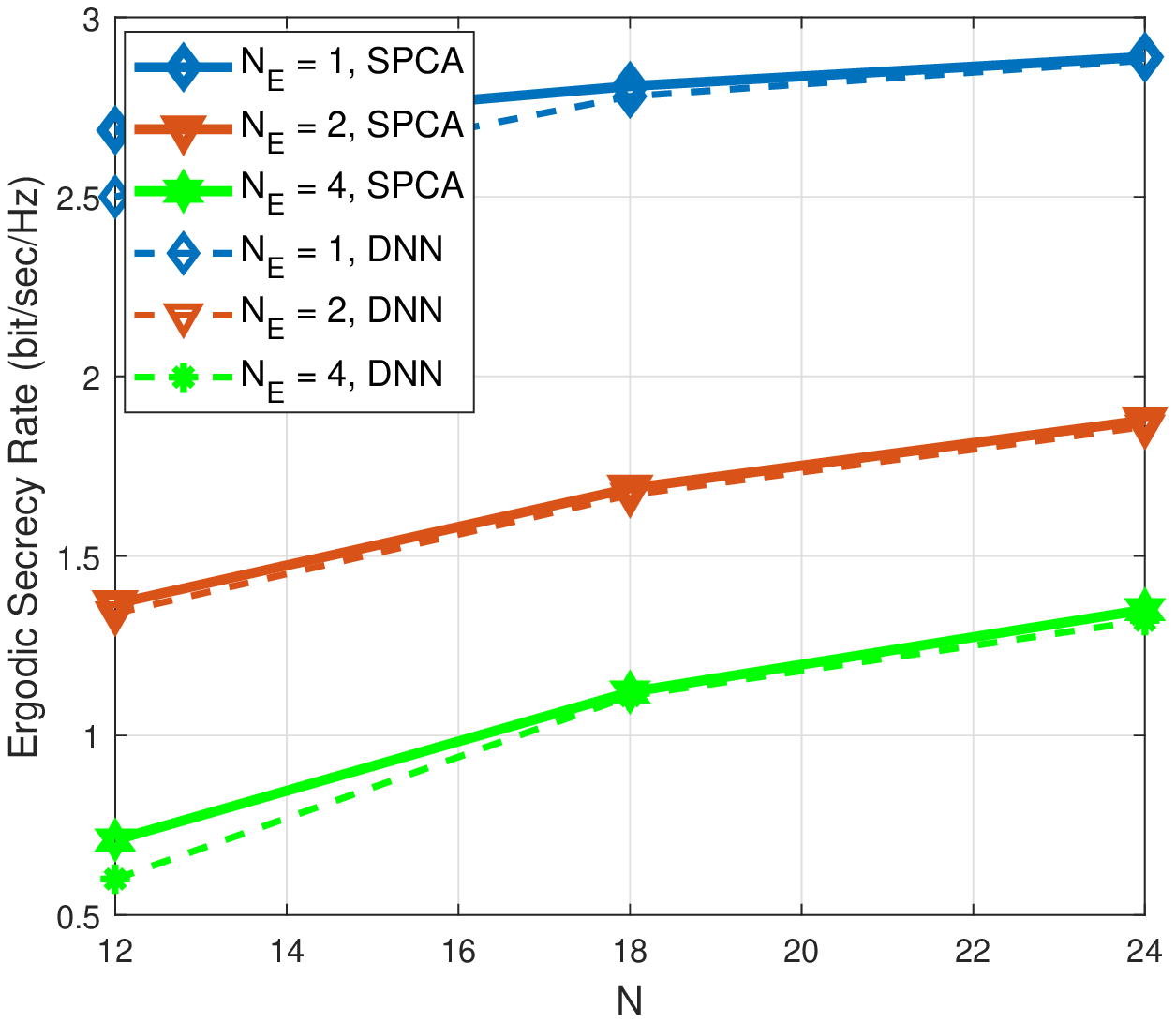}\caption{Ergodic secrecy rate of the proposed DNN scheme and the SPCA-based
scheme versus the total number of relay nodes $N$}
\par\end{center}%
\end{minipage}\textcolor{black}{~~}%
\begin{minipage}[t]{0.47\columnwidth}%
\begin{center}
\includegraphics[scale=0.68]{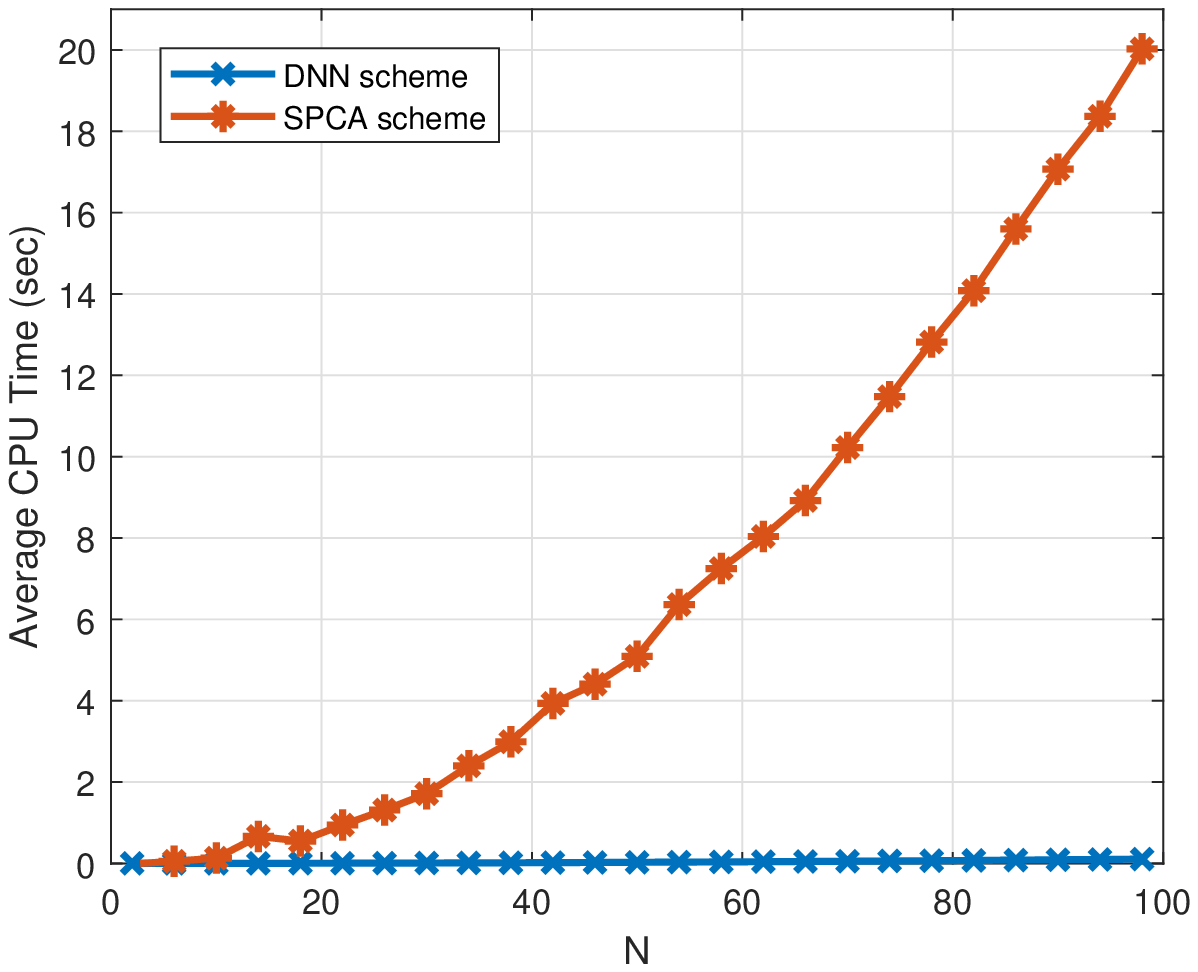}\caption{Average CPU time of the proposed DNN scheme and the sub-optimal scheme
(SPCA) versus the total number of relay nodes $N$}
\par\end{center}%
\end{minipage}
\end{figure}
 which there exists a low power to be allocated for jamming signals,
the artificial noise due to impairment plays the role of jamming.
As a consequence, upon increasing the impairment level, it is preferred
to allocate the major part of the power to the source and let the
inherent artificial noise due to impairment assist in safeguarding
the communication. Hence it is no longer expected to have lower secrecy
rate upon increasing the impairment levels at low power budgets. 

As mentioned before, the method presented in this paper was designed
for the case where we have perfect knowledge about external Eve. The
resultant maximization problem on secrecy rate, led to SPCA solution,
can be regarded as the upper bound. In contrast, if we assume no knowledge
about the external Eve (as already considered in \citep{8805054},
another secrecy scheme should be adopted, leading to a sub-optimal
solution which has considerably inferior performance than ours (as
it can be clearly seen in Fig.14). For a fair comparison, the minimum
QoS, $\gamma_{min}=12\:dB$, has been considered.

\textcolor{black}{Fig. 15 reveals the comparison result of our proposed
DNN-based scheme with SPCA-based scheme. We evaluate the ergodic screcy
rate versus total number of untrusted relay nodes $N$. Observing
Fig. 15, we can find that our DNN model provides a very accurate prediction
of the ASR , such that for $N=24$ the DNN-based scheme can achieve
99.61\% ASR performance of SPCA-based scheme. }

\textcolor{black}{As discussed before, DL can deal with the imposed
computational load upon increasing the network's dimensions, making
it an appropriate choice to satisfy the low-latency requirement of
B5G. In this respect, Fig. 16 shows the average CPU time versus the
number of relay nodes for the proposed DNN-based and SPCA-based schemes.
It can be clearly seen that, the DNN-based scheme requires the average
CPU time much less than SPCA-based scheme. For instance, }
\begin{figure}[tbh]
\centering{}\includegraphics[scale=0.6]{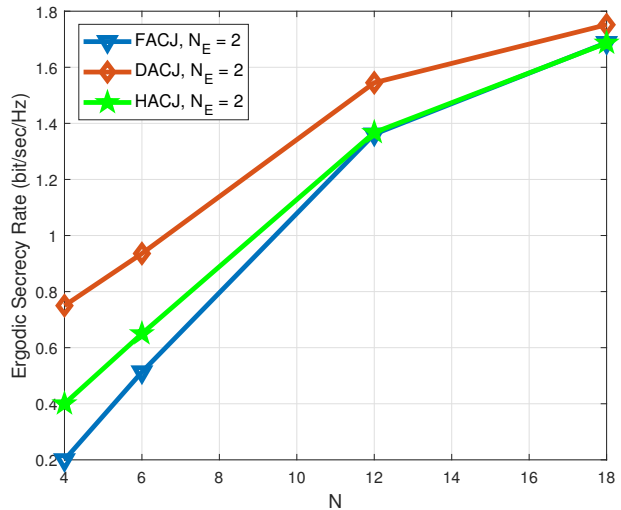}\caption{Ergodic secrecy rate of the proposed SPCA scheme versus the total
number of relay nodes N and different jammer selection scheme}
\end{figure}
\textcolor{black}{when total number of relay nodes power is 100, the
average CPU time of the proposed DNN-based scheme is 0.0031 sec but
that of SPCA-based is 20.934 sec, which is approximately 6,752.9 times.
Specifically, as number of relay nodes $N$ increases, the average
CPU time of the DNN-based scheme remains nearly constant but that
of SPCA-based grows exponentially due to an increasing number of iterations.}

Fig. 17 shows that unlike the assumption of unknown Eve's CSI in \citep{8805054},
which a relay selection algorithm, known as hybrid assisted cooperative
jamming (HACJ), was required to improve the secrecy, we can get rid
of the computational load imposed by relay selection using simple
DACJ, achieving the performance even better than HACJ. This is because,
assuming perfect CSIT and CSIR, instead of choosing a relay node as
a jammer the entire potential of relay nodes are preferred to forming
a centralized beam towards the legitimate destination whilst completely
nulling out the leakage at Eve. 

\section{Conclusions}

This paper has investigated PLS of an untrusted relaying network in
the face of realistic hardware impairment, where the source node communicates
in the presence of a multi-antenna Eve. The network relies on untrusted
relay nodes for increasing its communications quality, while aiming
for preserving the confidentiality of the information against the
combined eavesdropping attacks performed by both the untrusted relays
and a single Eves.We have assumed perfect CSIT and CSIR. This assumption
was indeed a strong one, hence the results represent the best-case
performance limit of practical relaying in the presence of HIs. In
this regard, the relay beamformer and the transmit powers were jointly
optimized to maximize the ASR under both the total and individual
power budget constraints of the entire network and each nodes, respectively.
Morever, DACJ was employed to safeguard the first cooperative phase.
On the other hand, for the second phase, the relay  beamformer was
adjusted for ensuring that the information leakage at Eve is entirely
removed. The resultant optimization problem was non-convex and solved
efficiently using the SPCA method. In order to prevent any failure
due to the solution's infeasibility, we have also proposed an iterative
initialization algorithm to find an initial point of the original
problem, leading to a feasible solution instead of relying on an arbitrary
point. Furthermore, to facilitate the implementation of proposed algorithm
in large-scale scenarios relying on numerous relays, a computationally
efficient data-driven approach was developed. A DL model was developed
to maximize the ASR performance, while the computational burden is
significantly reduced. Through extensive simulation results, we have
examined the effect of different system parameters on the ASR performance
as well as the efficiency of the proposed DL solution in large-scale
scenarios.

\end{document}